\documentclass[twocolumn,twocolappendix]{aastex63}

\newcommand{\carcsec}{\!\!\arcsec}
\newcommand{\ergs}{\mathrm{erg\ s^{-1}}}
\newcommand{\kms}{\mathrm{km\ s^{-1}}}
\newcommand{\Lya}{\mathrm{Ly}\alpha}
\newcommand{\TLya}{T^\mathrm{IGM}_{\Lya}}
\newcommand{\xHI}{x_\mathrm{HI}}
\newcommand{\fLya}{f_\mathrm{esc}^{\Lya}}
\newcommand{\fion}{f_\mathrm{esc}^\mathrm{ion}}

\received{January 13, 2021}
\revised{September 12, 2021; October 14, 2021}
\accepted{October 18, 2021}
\submitjournal{ApJ}

\shorttitle{Ly$\alpha$ luminosity function at $z=7.3$ and cosmic reionization}
\shortauthors{Goto et al.}

\begin{document}

\title{SILVERRUSH XI: Constraints on the Ly$\alpha$ luminosity function and cosmic reionization at $z=7.3$ with Subaru/Hyper Suprime-Cam}

\author{Hinako Goto}
\affiliation{Department of Astronomy, Graduate School of Science, The University of Tokyo, 7-3-1 Hongo, Bunkyo, Tokyo 113-0033, Japan}

\author[0000-0002-2597-2231]{Kazuhiro Shimasaku}
\affiliation{Department of Astronomy, Graduate School of Science, The University of Tokyo, 7-3-1 Hongo, Bunkyo, Tokyo 113-0033, Japan}
\affiliation{Research Center for the Early Universe, Graduate School of Science, The University of Tokyo, 7-3-1 Hongo, Bunkyo, Tokyo 113-0033, Japan}

\author{Satoshi Yamanaka}
\affiliation{Research Center for Space and Cosmic Evolution, Ehime University, 2-5 Bunkyo-cho, Matsuyama, Ehime 790-8577, Japan}
\affiliation{Waseda Research Institute for Science and Engineering, Faculty of Science and Engineering, Waseda University, 3-4-1, Okubo, Shinjuku, Tokyo 169-8555, Japan}

\author[0000-0002-8857-2905]{Rieko Momose}
\affiliation{Department of Astronomy, Graduate School of Science, The University of Tokyo, 7-3-1 Hongo, Bunkyo, Tokyo 113-0033, Japan}

\author{Makoto Ando}
\affiliation{Department of Astronomy, Graduate School of Science, The University of Tokyo, 7-3-1 Hongo, Bunkyo, Tokyo 113-0033, Japan}

\author{Yuichi Harikane}
\affiliation{Institute for Cosmic Ray Research, The University of Tokyo, 5-1-5 Kashiwanoha, Kashiwa, Chiba 277-8582, Japan}
\affiliation{Department of Physics and Astronomy, University College London, Gower Street, London WC1E 6BT, UK}

\author{Takuya Hashimoto}
\affiliation{Tomonaga Center for the History of the Universe (TCHoU), Faculty of Pure and Applied Sciences, University of Tsukuba, Tsukuba, Ibaraki 305-8571, Japan}

\author{Akio K. Inoue}
\affiliation{Department of Physics, School of Advanced Science and Engineering, Faculty of Science and Engineering, Waseda University, 3-4-1, Okubo, Shinjuku, Tokyo 169-8555, Japan}
\affiliation{Waseda Research Institute for Science and Engineering, Faculty of Science and Engineering, Waseda University, 3-4-1, Okubo, Shinjuku, Tokyo 169-8555, Japan}

\author{Masami Ouchi}
\affiliation{National Astronomical Observatory of Japan, 2-21-1 Osawa, Mitaka, Tokyo 181-8588, Japan}
\affiliation{Institute for Cosmic Ray Research, The University of Tokyo, 5-1-5 Kashiwanoha, Kashiwa, Chiba 277-8582, Japan}
\affiliation{Kavli Institute for the Physics and Mathematics of the Universe (Kavli IPMU, WPI), The University of Tokyo, 5-1-5 Kashiwanoha, Kashiwa, Chiba, 277-8583, Japan}

\begin{abstract}
The Ly$\alpha$ luminosity function (LF) of Ly$\alpha$ emitters (LAEs) has been used to constrain the neutral hydrogen fraction in the intergalactic medium (IGM) and thus the timeline of cosmic reionization.
Here we present the results of a new narrow-band imaging survey for $z=7.3$ LAEs in a large area of $\sim 3\ \mathrm{deg}^2$ with Subaru/Hyper Suprime-Cam.
No LAEs are detected down to $L_{\Lya}\simeq 10^{43.2}\ \ergs$ in an effective cosmic volume of $\sim 2\times 10^6$ Mpc$^3$, placing an upper limit to the bright part of the $z=7.3$ Ly$\alpha$ LF for the first time and confirming a decrease in bright LAEs from $z=7.0$.
By comparing this upper limit with the Ly$\alpha$ LF in the case of the fully ionized IGM, which is predicted using an observed $z=5.7$ Ly$\alpha$ LF on the assumption that the intrinsic Ly$\alpha$ LF evolves in the same way as the UV LF, we obtain the relative IGM transmission $\TLya(7.3)/\TLya(5.7)<0.77$, and then the volume-averaged neutral fraction $\xHI(7.3)>0.28$. 
Cosmic reionization is thus still ongoing at $z=7.3$, being consistent with results from other $\xHI$ estimation methods.
A similar analysis using literature Ly$\alpha$ LFs finds that at $z=6.6$ and 7.0 the observed Ly$\alpha$ LF agrees with the predicted one, consistent with full ionization.
\end{abstract}


\section{Introduction}
\label{sec:intro}
Cosmic reionization is a key process in the early universe where massive stars and/or active galactic nuclei ionized the intergalactic medium (IGM) hydrogen that had been neutral after recombination at redshift ($z$) $\sim1100$.
Understanding how and when this process occurred is one of the major goals of modern cosmology and astronomy.

The Thomson scattering optical depth of the cosmic microwave background (CMB) suggests a mid-point reionization redshift of $z_\mathrm{re}=7.7\pm0.7$ \citep{Planck2020}. 
Furthermore, recent observations of various kinds of distant objects beyond $z\sim6$ have constrained the period of reionization, by estimating the neutral hydrogen fraction in the IGM, $\xHI$, as a function of redshift.
Gunn-Peterson troughs in quasi-stellar object (QSO) spectra suggest that cosmic reionization has been completed by $z\sim6$ \citep[e.g.,][]{Fan2006}.
Damping wing signatures in QSO spectra \citep[e.g.,][]{Schroeder2013, Greig2017, Greig2019, Banados2018, Davies2018, Wang2020} and gamma-ray burst (GRB) spectra \citep[e.g.,][]{Totani2006, Totani2014} have also placed constraints on $\xHI$ at $z\simeq6-7.5$ although the total number of observed sources is very limited.

Ly$\alpha$ emission from galaxies is also a powerful probe of $\xHI$ because galaxies are much more numerous than QSOs and GRBs. Methods using galaxies' Ly$\alpha$ emission include:
the fraction of Lyman break galaxies (LBGs) that emit Ly$\alpha$ \citep[e.g.,][]{Stark2011,Ono2012,Mesinger2015}, the Ly$\alpha$ equivalent-width (EW) distribution of LBGs \citep[e.g.,][]{Mason2018a, Mason2019, Hoag2019, Whitler2020, Jung2020}, and the Ly$\alpha$ luminosity function (LF) of Ly$\alpha$ emitters (LAEs; i.e., galaxies with strong Ly$\alpha$ emission; e.g.,  \citealt{Kashikawa2006, Kashikawa2011, Ouchi2010, Konno2014, Konno2018, Zheng2017, Ota2017, Itoh2018, Hu2019}).

The observations mentioned above collectively suggest that the universe is significantly neutral at $z \gtrsim 7$. However, the $\xHI$ estimates at $z>7$ still have a large scatter, spanning $\xHI \sim 0.2-0.9$, perhaps suggesting field-to-field variation or the presence of a systematic uncertainty in each method.
To further constrain the time evolution of $\xHI$, we need to accumulate estimates by individual methods.

In this study, we focus on the Ly$\alpha$ LF method. This method estimates $\xHI$ by comparing an observed Ly$\alpha$ LF of LAEs at a target redshift with that after completion of reionization (e.g., $z=5.7$). 
This method's advantage is that $\xHI$ can be estimated with a negligibly small redshift uncertainty for a large cosmic volume  if an LAE sample from a large-area narrow-band (NB) survey is used.
A drawback is that the effect of galaxy evolution on observed LFs has to be eliminated using the UV LF of LBGs or a theoretical model.

The Ly$\alpha$ LF in the reionization era has been obtained at $z=5.7$, 6.6, 7.0, and 7.3.
At $z=7.3$, the highest redshift that can be probed with optical CCD detectors, \cite{Konno2014} have obtained $\xHI=0.3-0.8$ from an accelerated decline of the Ly$\alpha$ LF from $z=5.7$.
However, because of a relatively small survey area, they have derived only the faint ($L_{\Lya} < 10^{42.9}\ \ergs$) part of the Ly$\alpha$ LF. 
This is contrasted to the studies at lower redshifts that cover up to $L_{\Lya} \gtrsim 10^{43.5}\ \ergs$ thanks to a large survey volume of $>1\times 10^6$ Mpc$^3$.
To obtain a robust $\xHI$ estimate at $z=7.3$, we need to compare the entire LF including the bright part with that at $z=5.7$. 

In this paper, we present the results of a new survey of bright $z=7.3$ LAEs with Subaru/Hyper Suprime-Cam (HSC; \citealt{Miyazaki2012, Miyazaki2018, Komiyama2018, Kawanomoto2018, Furusawa2018}), conducted as part of the SILVERRUSH project (\citealt{Ouchi2018, Shibuya2018a, Shibuya2018b, Konno2018, Harikane2018, Inoue2018, Higuchi2019, Harikane2019, Kakuma2019}; Ono et al. in prep.) that uses four HSC NB filters to study LAEs.
Since our survey volume is as large as $\sim 2\times 10^6$ Mpc$^3$, which is seven times larger than \cite{Konno2014}'s, our constraint on $\xHI$ should also be robust against the uncertainty due to spatially inhomogeneous reionization (see Section \ref{sec:images}).
To infer $\xHI$, we first need to calculate the Ly$\alpha$ transmission of the IGM, $\TLya$. 
We use a new method to calculate $\TLya$ and apply it also to the previous studies' Ly$\alpha$ LFs at $z=6.6$ \citep{Konno2018}, 7.0 (\citealt{Itoh2018}; \citealt{Hu2019}), and 7.3 \citep{Konno2014}.

This paper is structured as follows. Section 2 describes the HSC imaging data used in this study. Section 3 describes our LAE selection. Section 4 presents new constraints on the Ly$\alpha$ LF, $\TLya$, and $\xHI$, including a comparison with $\xHI$ estimates by other methods. 
Section 5 is devoted to conclusions.

Throughout this paper, we assume a flat $\Lambda$CDM cosmology with $\Omega_m=0.3$, $\Omega_\Lambda=0.7$, and $H_0=70\ \mathrm{km\ s^{-1}\ Mpc^{-1}}$.
Magnitudes are given in the AB system \citep{Oke1983}. 
Distances are in comoving units unless otherwise noted.

\begin{deluxetable*}{ccccccc}
\tablecaption{Summary of Imaging Data\label{tab:data}}
\tablehead{\colhead{Field} & \colhead{Band} & \colhead{Area$^*$} & \colhead{Exposure Time} & \colhead{PSF Size$^\dag$} & \colhead{$m_{{\rm lim}}$$^{\dag\ddag}$} & \colhead{Dates of Observation}\\
&& \colhead{(deg$^2$)} & \colhead{(s)} & \colhead{(arcsec)} & \colhead{(mag)} &}
\startdata
COSMOS & NB1010 & 1.55 & 49,122 & 0.69 & 24.7 & 2018 Feb 12, 13, 21, 2019 Jan 5\\
& $y$ &&& 0.70 & 26.4 &\\
& $z$ &&& 0.59 & 27.1 &\\
& $i$ &&& 0.64 & 27.5 &\\
& $r$ &&& 0.67 & 27.7 &\\
& $g$ &&& 0.80 & 28.2 &\\
& NB921 &&& 0.64 & 26.4 &\\
& NB816 &&& 0.63 & 26.5 &\\ \hline
SXDS & NB1010 & 1.47 & 50,400 & 0.72 & 24.5 & 2018 Feb 12, 13, 21, 2019 Jan 4, 5, 9\\
& $y$ &&& 0.58 & 25.5 &\\
& $z$ &&& 0.56 & 26.3 &\\
& $i$ &&& 0.62 & 27.0 &\\
& $r$ &&& 0.62 & 27.1 &\\
& $g$ &&& 0.66 & 27.6 &\\
& NB921 &&& 0.69 & 26.2 &\\
& NB816 &&& 0.58 & 26.3 &
\enddata
\tablenotetext{*}{The effective survey area  after removing masked regions (Section \ref{sec:images} and Figure \ref{fig:surveyarea}).}
\tablenotetext{\dag}{Values calculated at {\tt patches} in the central region of the field of view, 9813-5,4 ({\tt tract-patch}) in COSMOS and 8523-2,6 in SXDS (see Figure \ref{fig:limitmag}).}
\tablenotetext{\ddag}{The $5\sigma$ limiting magnitude in an aperture with a diameter of two times the PSF FWHM, $1.\carcsec41$ (COSMOS) and $1.\carcsec44$ (SXDS).}
\end{deluxetable*}

\section{Data}
In this work, we use the images in the HSC Subaru Strategic Program\footnote{The SILVERRUSH project uses the data from this program.} (SSP; \citealt{Aihara2018}) S19A release data, which are reduced with {\tt hscPipe 7} \citep{Bosch2018}.\footnote{\url{https://hsc.mtk.nao.ac.jp/pipedoc/pipedoc_7_e/index.html}}
To search for LAEs at $z=7.3$, we use the custom-made narrow-band filter NB1010.

\subsection{NB1010 Filter}
\label{sec:NB1010}
\begin{figure}[tb!]
    \centering
    \includegraphics[width=\linewidth]{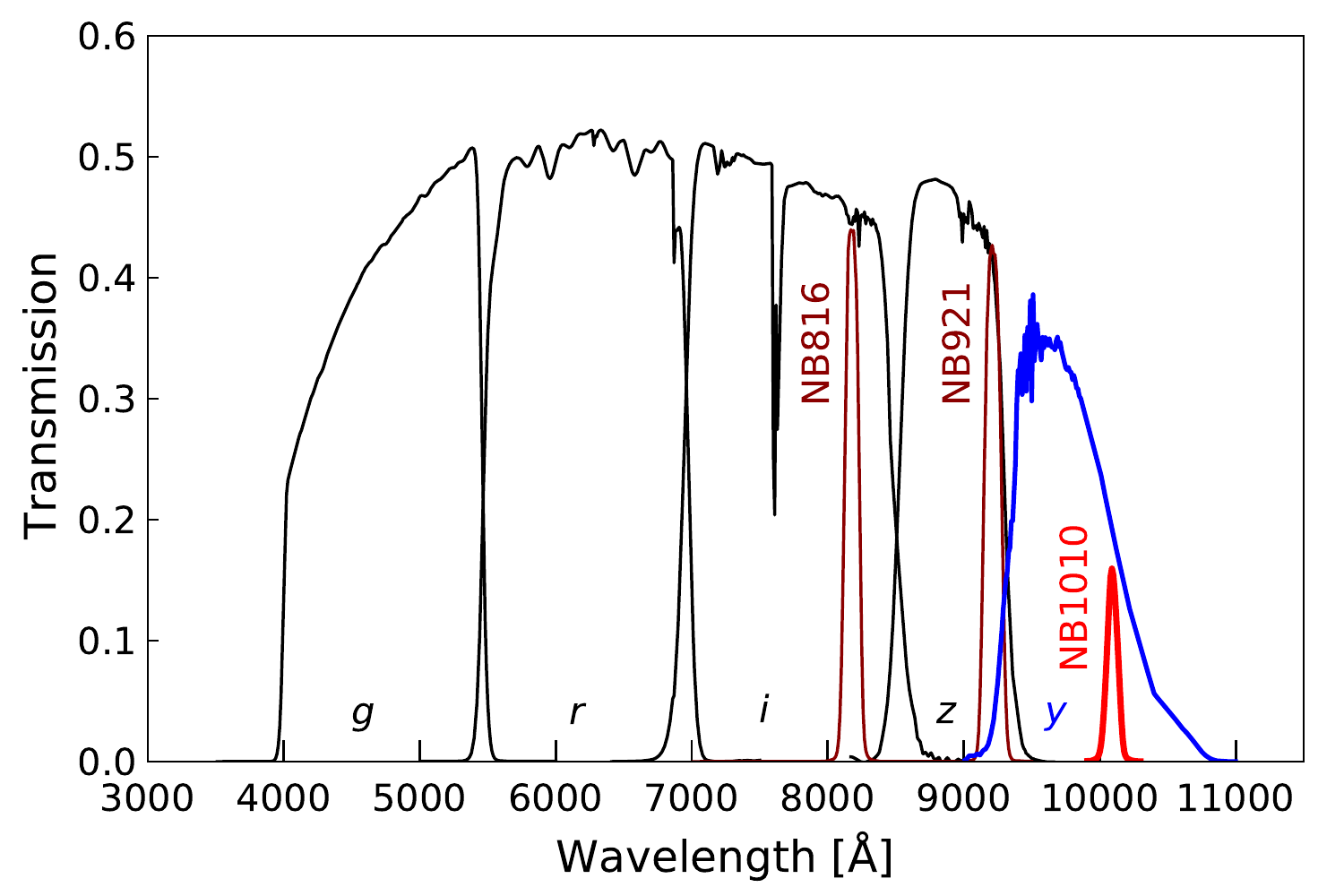}
    \caption{Total transmission curves of the HSC NB and BB filters used in this study, including the CCD quantum efficiency; the transmission of the dewar window, the primary focus unit, and the atmosphere; and the reflectivity of the primary mirror.}
    \label{fig:transmission}
\end{figure}

The NB1010 filter has a central wavelength of $\lambda_c = 10092$ \AA\ and an FWHM of 91 \AA\ to identify the Ly$\alpha$ emission at $z=7.30\ \pm\ 0.04$.
At a given observing time, a filter with a narrower FWHM can detect fainter Ly$\alpha$ emission.
With the interference coating technique when this filter was manufactured, the above FWHM value was practically the narrowest that could be achieved around 10000 \AA\ in the very fast (F/2.25) incoming beam of the HSC.
Figure \ref{fig:transmission} shows the transmission curves of the NB and broad-band (BB) filters used in this study (NB1010, $y$, $z$, $i$, $r$, $g$, NB921, and NB816).

\begin{figure*}[tb!]
    \centering
    \includegraphics[width=0.9\linewidth]{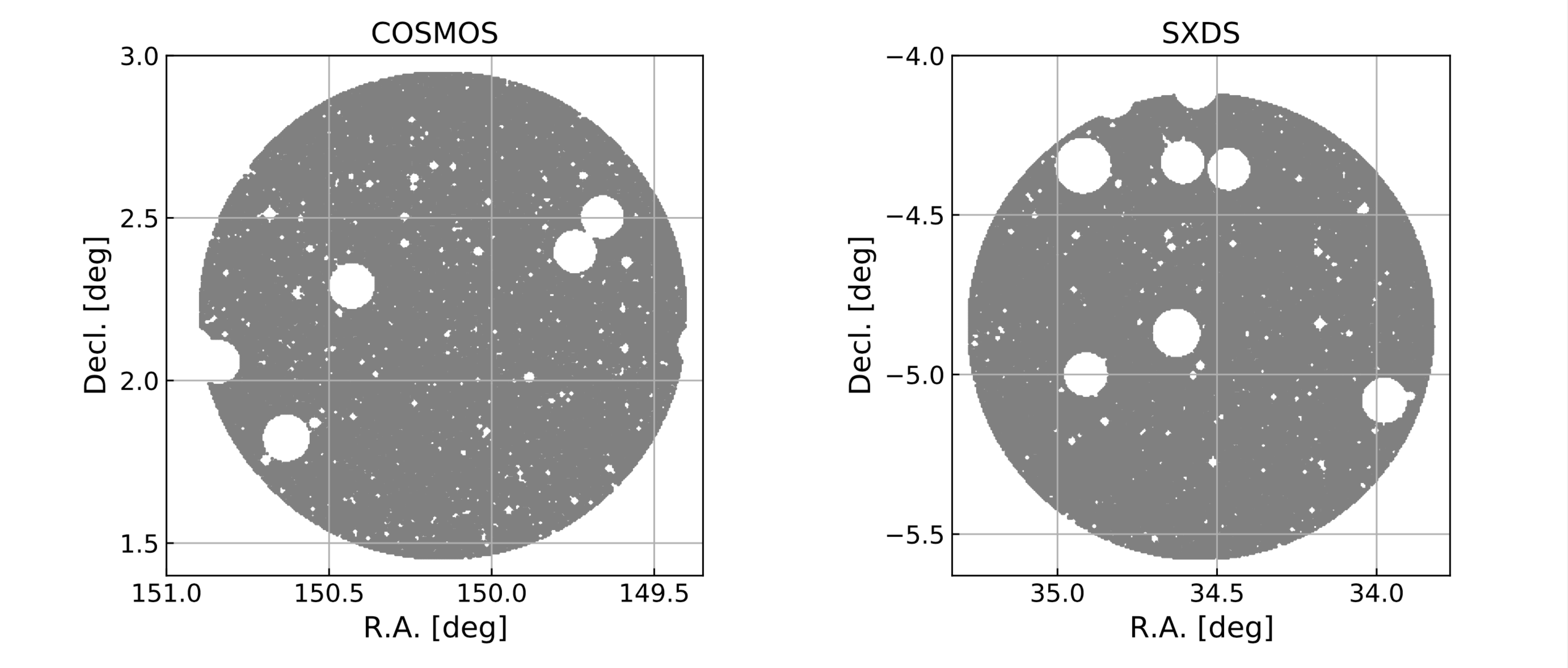}
    \caption{Effective survey areas (gray-filled regions) in the COSMOS and SXDS fields. Masked regions are shown in white.}
    \label{fig:surveyarea}
\end{figure*}

\begin{figure*}[tb!]
    \centering
    \includegraphics[width=0.9\linewidth]{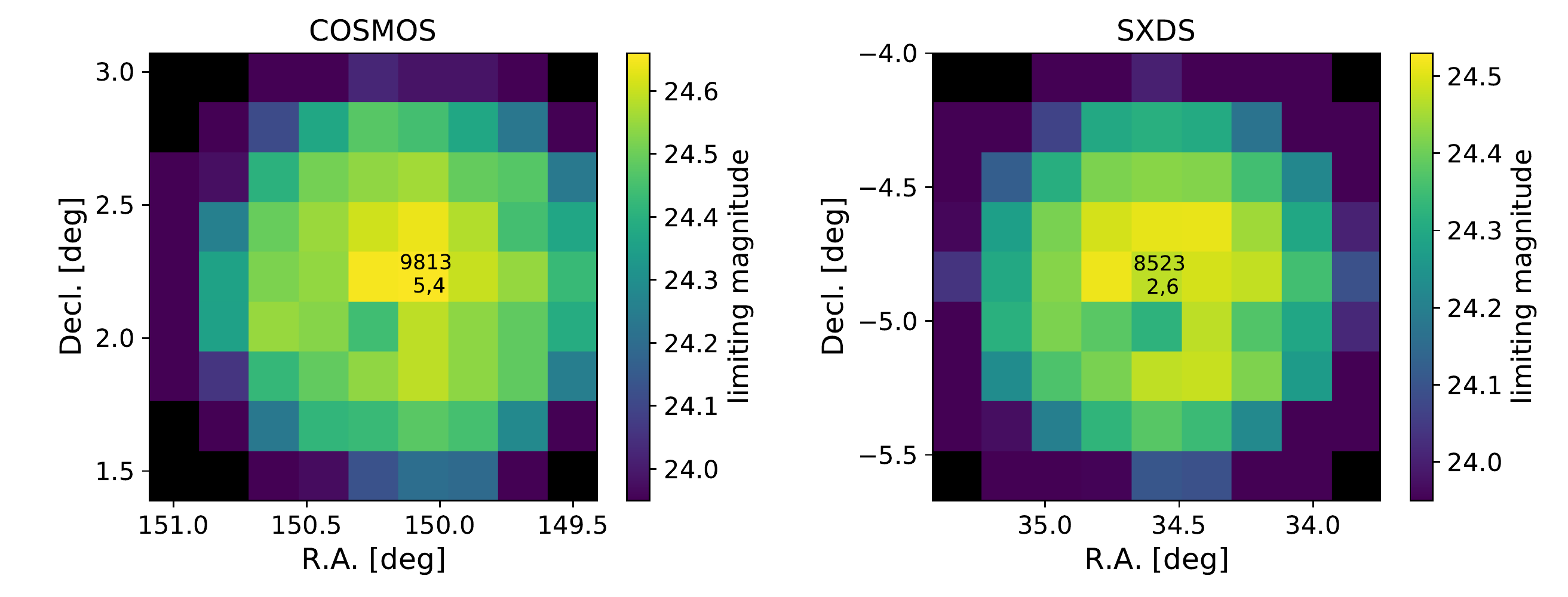}
    \caption{$5\sigma$ limiting magnitude maps of the NB1010 images in the two fields. Limiting magnitudes are measured in an aperture with a diameter of two times the PSF FWHM,
    $1.\carcsec41$ (COSMOS) and $1.\carcsec44$ (SXDS).
    Each square represents a {\tt patch}, and the number at the center of each image represents the number of {\tt tract} and {\tt patch}, for example [{\tt tract} = 9813, {\tt patch} = 5,4] for COSMOS.}
    \label{fig:limitmag}
\end{figure*}

\subsection{Images}
\label{sec:images}
The details of the NB and BB imaging data used in this study are summarized in Table \ref{tab:data}.
The NB1010 observations were carried out between 2018 February and 2019 January in two fields, COSMOS and SXDS.
The total exposure time is 13.6 hr in COSMOS and 14.0 hr in SXDS, respectively.

To mask out regions around bright stars, we use the mask images provided by the pipeline.
In our analysis, we do not use pixels which have either the {\tt SAT}, {\tt BRIGHT\_OBJECT}, or {\tt NO\_DATA} flag.\footnote{As for {\tt BRIGHT\_OBJECT}, we use the S18A mask images instead, because the S19A mask images do not have this flag.}
We also remove low signal-to-noise ratio (S/N) regions near the edges of the images.
After removal of these masked regions, the effective survey areas are $1.55\ {\rm deg^2}$ and $1.47\ {\rm deg^2}$ in the COSMOS and SXDS fields, respectively.
Figure \ref{fig:surveyarea} shows the effective survey areas of the two fields.
Assuming a top-hat NB filter with the FWHM of NB1010, the survey volumes are $8.79 \times 10^5\ {\rm Mpc^3}$ and $8.35 \times 10^5\ {\rm Mpc^3}$ in COSMOS and SXDS, respectively.
At $z=7.3$ and with $\xHI \sim 0.1$, the volume of typical ionized bubbles is estimated to be $\sim 3 \times 10^5 \ \mathrm{Mpc}^3$ using an analytic model by \cite{Furlanetto2005}.
Since our total survey volume is $\sim 5$ times larger than this, our constraint on $\xHI$ should be robust against the uncertainty due to spatially inhomogeneous reionization.

In the HSC-SSP data processing, the sky is divided into grids called {\tt tracts}, and each {\tt tract} is further divided into sub-areas called {\tt patches}.
Each {\tt patch} covers approximately $12\arcmin \times 12\arcmin$ of the sky \citep{Aihara2018}.
We conduct LAE selection at each {\tt patch}, using the local limiting magnitude estimated at that {\tt patch}.
To do so, we estimate limiting magnitudes at all {\tt patches}, as shown in Figure \ref{fig:limitmag}, by placing in the unmasked region random apertures whose diameter is two times the point-spread function (PSF) FWHM averaged over the image, $1.\carcsec41$ (COSMOS) and $1.\carcsec44$ (SXDS).
For each field, the limiting magnitude gradually becomes brighter toward the edge of the image.
At {\tt patches} in the central region, the NB1010 images have seeing sizes of $0.\carcsec69$ (COSMOS) and $0.\carcsec72$ (SXDS), and reach $5\sigma$ limiting magnitudes of 24.7 mag (COSMOS) and 24.5 mag (SXDS).

\section{LAE Selection}
\subsection{Source Detection and Photometry}
\label{sec:SExtractor}
We use {\tt SExtractor} version 2.19.5 \citep{Bertin1996} for source detection and photometry.
Object detection is first made in the NB1010 images, and photometry is then performed in the other band images using the double-image mode.
We set {\tt SExtractor} configuration parameters so that an area equal to or larger than 3 contiguous pixels with a flux greater than 1.5$\sigma$ of the background RMS is considered as a separate object.
An aperture magnitude, {\tt MAG\_APER}, is measured with an aperture size of two times the PSF FWHM, $1.\carcsec41$ (COSMOS) and $1.\carcsec44$ (SXDS), and used for the LAE selection (Section \ref{sec:LAEselection}).
Magnitudes and colors are corrected for Galactic extinction using \cite{Schlegel1998}.

\subsection{LAE Selection}
\label{sec:LAEselection}
We select $z=7.3$ LAE candidates based on
(1) significant detection in the NB1010 image,
(2) NB color excess due to the Ly$\alpha$ emission, $y-{\rm NB1010}$, and
(3) no detection in the bluer bands to exclude foreground galaxies.
The exact selection criteria are as follows:
\begin{eqnarray}
&{\rm NB1010} <{\rm NB1010}_{5\sigma}, \nonumber\\
&y-{\rm NB1010}>1.9, \nonumber\\
&z>z_{3\sigma},\ i>i_{3\sigma},\ r>r_{3\sigma},\ g>g_{3\sigma}, \nonumber\\
&{\rm NB921}>{\rm NB921}_{3\sigma},\ {\rm NB816}>{\rm NB816}_{3\sigma},
\label{eq:criteria}
\end{eqnarray}
where ${\rm NB1010}_{5\sigma}$ is the 5$\sigma$ limiting magnitude of NB1010,
and [$z_{3\sigma}$, $i_{3\sigma}$, $r_{3\sigma}$, $g_{3\sigma}$, ${\rm NB921}_{3\sigma}$, and ${\rm NB816}_{3\sigma}$] are the 3$\sigma$ limiting magnitudes of [$z$, $i$, $r$, $g$, ${\rm NB921}$, and ${\rm NB816}$] bands.
Note that we use the limiting magnitude estimated at the {\tt patch} in which the object exists (see Section \ref{sec:images}).
We use aperture magnitudes, {\tt MAG\_APER}, (Section \ref{sec:SExtractor}) to measure S/Ns and colors.
To measure colors accurately, we convolve the $y$ image of the SXDS field to have the same PSF size as the NB1010 image.

\begin{figure}[tb!]
    \centering
    \includegraphics[width=\linewidth]{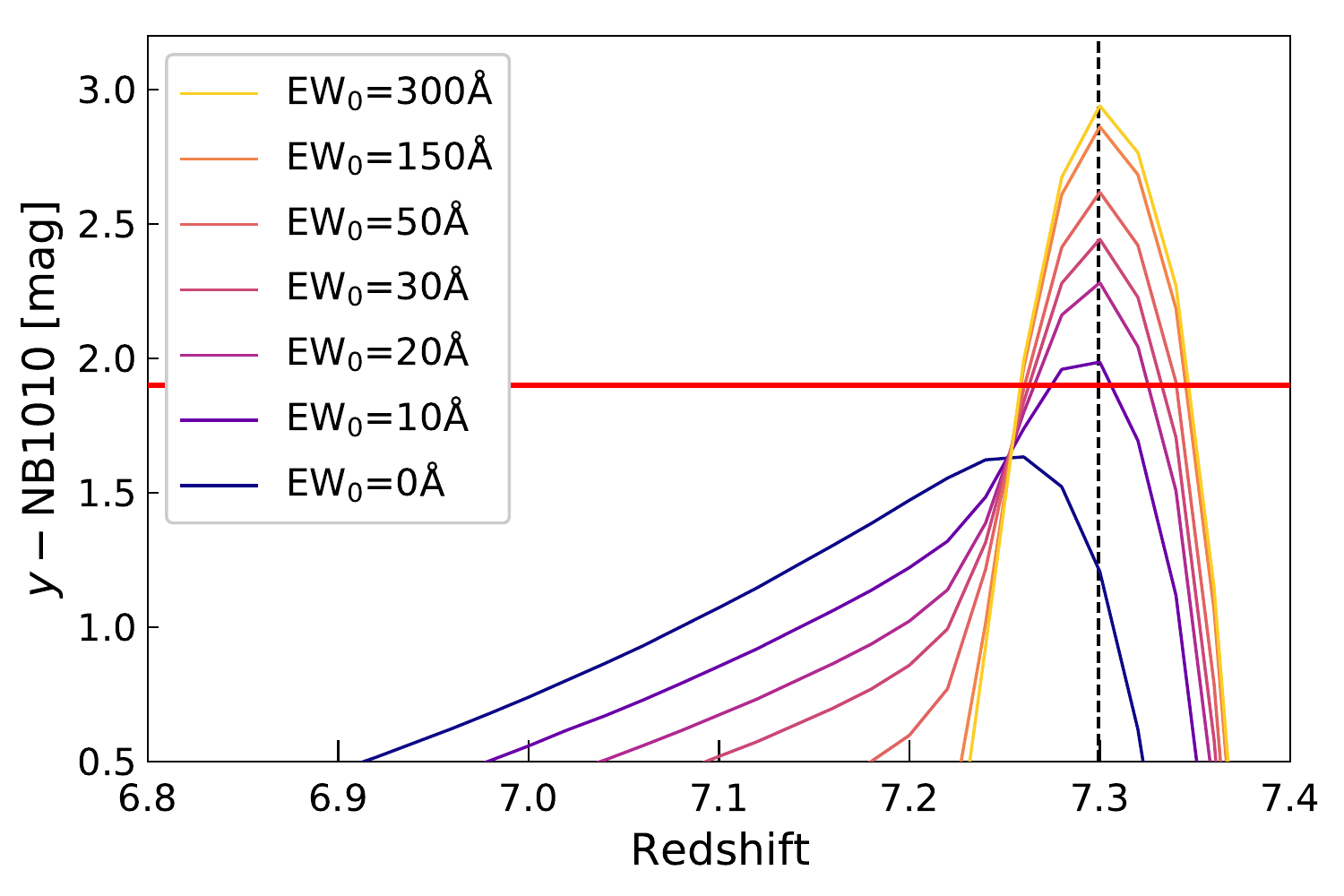}
    \caption{Expected $y-$NB1010 colors of our model LAEs as a function of redshift. The black dashed line shows our target redshift, $z=7.3$. The red horizontal line shows the color criterion we adopt, $y-\mathrm{NB1010}>1.9$.}
    \label{fig:modelgal}
\end{figure}

To determine the $y-$NB1010 color criterion above, we calculate the expected colors of $z=7.3$ LAEs.
We assume a simple model spectrum that has a flat continuum ($f_\nu =$ const., i.e., the UV continuum slope $\beta=-2$)\footnote{\cite{Konno2014}, \cite{Itoh2018}, \cite{Konno2018}, and \cite{Hu2019} have also adopted $\beta=-2$. Besides, \cite{Itoh2018} have found that $\beta=0, -1, -2$, and $-3$ give similar results.}
and $\delta$-function Ly$\alpha$ emission with rest-frame equivalent widths of EW$_0$ = 0, 10, 20, 30, 50, 150, and 300 \AA.
Then we redshift the spectra and apply IGM absorption \citep{Madau1995}.\footnote{Because the transmittance at wavelengths shorter than Ly$\alpha$ is almost zero, adopting a different model (e.g., \citealt{Inoue2014}) does not change our color criterion.}
The colors of the spectra are calculated with the transmission curves of the HSC filters shown in Figure \ref{fig:transmission}.
Figure \ref{fig:modelgal} shows the results of the expected colors as a function of redshift.
Based on Figure \ref{fig:modelgal}, we adopt $y-\mathrm{NB1010}>1.9$ as our color criterion for $z=7.3$ LAEs, corresponding to $\mathrm{EW}_0 \gtrsim 10$ \AA.
Since we want to see the evolution from $z=5.7$ to estimate $\xHI$, we adopt the same EW limit as \cite{Konno2018}'s $z=5.7$ LAEs.
This EW limit is also the same as those adopted in \cite{Konno2018} for $z=6.6$ LAEs and \cite{Itoh2018} and \cite{Hu2019} for $z=7.0$ LAEs.

We apply the selection criteria to the objects detected in Section \ref{sec:SExtractor}.
Then we perform visual inspection of the objects that pass the selection criteria.
Spurious sources such as cosmic rays, CCD artifacts, and artificial diffuse objects outside the masked regions are removed.
Example images of the spurious sources are shown in Appendix \ref{appendix:A}.
After the visual inspection, there are no LAE candidates left in either the COSMOS or SXDS field.

\subsection{Sample Incompleteness}
\label{sec:comp}
To estimate what fraction of real LAEs pass our selection, we insert pseudo-LAEs into the NB1010 image of each field, and then calculate ``detection completeness'' and ``selection completeness''.

\subsubsection{pseudo-LAEs}
\label{sec:pseudoLAE}
We use {\tt GALSIM} \citep{Rowe2015} to simulate pseudo-LAEs.
The pseudo-LAEs have a S\'{e}rsic index of $n=1.0$, and a half-light radius of $r_e\sim 0.8\ \mathrm{kpc}$ (physical units), which corresponds to $0.\carcsec16$ at $z=7.3$.
These values are consistent with those of $z\sim 7$ LBGs (e.g., \citealt{Shibuya2015}; \citealt{Kawamata2018}) at $M_\mathrm{UV} \lesssim -21$, corresponding to the luminosity limit of this study, $\log L_{\Lya}\ [\ergs] \gtrsim 43.2$, and typical rest-frame Ly$\alpha$ EWs at this redshift, $\mathrm{EW}_0\lesssim 100$ \AA\ (e.g., \citealt{Hashimoto2019}).
Previous studies have also adopted similar values (\citealt{Itoh2018}; \citealt{Konno2018}; \citealt{Hu2019}).

Most LAEs have an extended Ly$\alpha$ halo component \citep[e.g.,][]{Momose2016, Leclercq2017} that cannot be detected in NB images.
\cite{Hu2019} have simulated pseudo-LAEs with larger half-light radii of 0.9, 1.2, and 1.5 kpc (physical units), taking account of the total Ly$\alpha$ emission (main body plus halo) based on MUSE observations of $z=3$--$6$ LAEs by \cite{Leclercq2017}, and found negligible differences in the completeness measurements with these radii.
Since the sizes and luminosities of Ly$\alpha$ halo components at $z>6$ are yet to be examined, we adopt $r_e\sim 0.8\ \mathrm{kpc}$ (physical units) that is consistent with the values assumed in the previous studies.
We apply PSF convolution to the pseudo-LAEs and randomly insert them into the NB1010 images avoiding the masked regions.

\cite{Drake2017} have shown with MUSE data of LAEs over $3\lesssim z \lesssim 6$ that extended Ly$\alpha$ emission affects the detection completeness of LAEs.
However, if the ratio of the extended component to the total luminosity does not evolve with redshift, all NB surveys will be underestimating the total luminosity and incompleteness in the same way, which does not affect $\xHI$ estimates.
Indeed, Figure 4 of \cite{Drake2017} shows that the contribution of the extended component is almost the same regardless of redshift and brightness.

\subsubsection{Detection Completeness and Selection Completeness}
\label{sec:det_sel}
We perform source detection and photometry for the pseudo-LAEs with {\tt SExtractor} in exactly the same manner as in Section \ref{sec:SExtractor} to calculate detection completeness.
We define detection completeness as the fraction in number of detected pseudo-LAEs to the input pseudo-LAEs.

We also calculate selection completeness, which has been introduced in \cite{Hu2019}, to account for the effects of not meeting the LAE selection criteria because of foreground sources.
Some LAEs may be blended with foreground sources that are not bright enough in NB1010 to hinder the LAEs' detection but bright enough in $y$ or bluer bands to prevent them from passing the selection defined as lines 2--4 of Equation (\ref{eq:criteria}).
To calculate this selection completeness, we assume underlying broadband fluxes to be zero and insert pseudo-LAEs only into the NB1010 images, following \cite{Hu2019} (see Section 4.1 of \citealt{Hu2019} for more details).
Selection completeness is defined as the fraction in number of pseudo-LAEs which meet the selection criteria (lines 2--4 of Equation (\ref{eq:criteria})) to the detected pseudo-LAEs.

\begin{figure}[tb!]
    \centering
    \includegraphics[width=\linewidth]{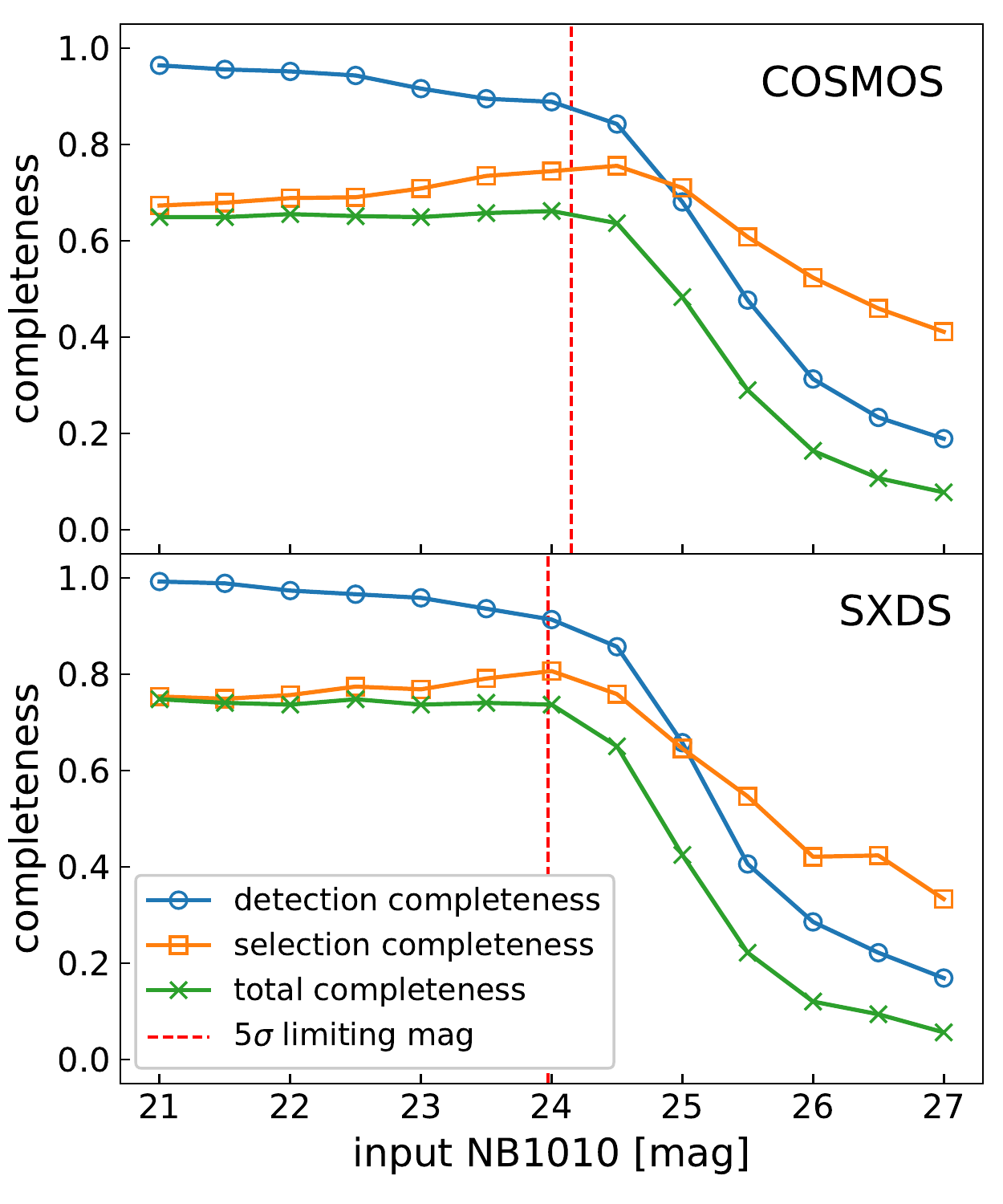}
    \caption{Detection completeness, selection completeness, and total completeness (the product of detection completeness and selection completeness) as a function of input NB1010 total magnitude for 9813-4,4 (COSMOS; top) and 8523-2,6 (SXDS; bottom)  {\tt patches} in the central regions. The red dashed lines denote the 5$\sigma$ limiting magnitude for each {\tt patch}, corrected for the offset between the input total magnitude and {\tt MAG\_APER}.}
    \label{fig:comp}
\end{figure}

As an example, Figure \ref{fig:comp} shows the results for 9813-4,4 (COSMOS) and 8523-2,6 (SXDS) {\tt patches} in the central regions.
Their detection completeness and selection completeness are $\gtrsim 90\%$ and $\sim 70-80\%$, respectively, at magnitudes brighter than the $5\sigma$ limiting magnitude.
Note that selection completeness, which has not been considered in previous studies except in \cite{Hu2019}, is more dominant (i.e., lower) than detection completeness at input magnitudes $\lesssim 25$ mag.
Our results of selection completeness, 
$\sim 70-80\%$ at magnitudes brighter than the $5\sigma$ magnitude, 
are similar to those of \cite{Hu2019} and are mainly due to foreground contamination in bluer bands.
\cite{Hu2019} have estimated the effect of blending with foreground sources in bluer bands by random aperture photometry.
For example, they have found that only a $74.9\%$ area of the COSMOS HSC $g$-band image has S/N$<3\sigma$ with an aperture size of $1.\carcsec35$.

As found in Figure \ref{fig:comp}, selection completeness has a mild peak near the $5\sigma$ limiting magnitude.
As explained by \cite{Hu2019}, fainter LAEs would not be detected in the detection image (NB1010 in our case) if blended with a foreground source, which results in a lower detection completeness.
Consequently, selection completeness gradually increases toward fainter magnitude because non-detected pseudo-LAEs, blended with a foreground source, are pre-excluded from the calculation, i.e., most of the detected pseudo-LAEs are located in sparse regions.
On the other hand, selection completeness drops at magnitudes fainter than the $5\sigma$ magnitude, as most of these LAEs are detected simply because they happen to be injected on top of a foreground source.

Note that we only consider detection completeness when calculating upper limits of the cumulative Ly$\alpha$ LF (Section \ref{sec:cumLF}) to directly compare them with the Ly$\alpha$ LF measurements by previous studies which have not considered selection completeness.
We use the detection completeness averaged over the effective area, 0.95 (COSMOS) and 0.96 (SXDS).

\section{Results and Discussion}
\subsection{Cumulative Ly$\alpha$ Luminosity Function}
\label{sec:cumLF}
From the result of no detection of $z=7.3$ LAEs (Section \ref{sec:LAEselection}), we calculate upper limits of the cumulative Ly$\alpha$ LF.
The upper limit of the cumulative number density at a given $L_{\Lya}$ is calculated as:
\begin{eqnarray}
n(>L_{\Lya}) < \frac{1.15}{V_{\mathrm{eff}, 1} f_{\mathrm{det}, 1} + V_{\mathrm{eff}, 2} f_{\mathrm{det}, 2}}, 
\label{eq:upperlim}
\end{eqnarray}
where 1.15 corresponds to the $68 \%$ upper limit for no detection assuming the Poisson statistics,
$V_\mathrm{eff}$ is the total survey volume of {\tt patches} whose limiting luminosity is fainter than this $L_{\Lya}$ (i.e., {\tt patches} that allow LAE search at $> L_{\Lya}$), 
and $f_\mathrm{det}$ is the detection completeness derived in Section \ref{sec:det_sel}, 0.95 (COSMOS) and 0.96 (SXDS).
The subscripts 1 and 2 in Equation (\ref{eq:upperlim}) represent the COSMOS and SXDS fields, respectively.

\begin{figure*}[htb!]
    \centering
    \includegraphics[width=0.8\linewidth]{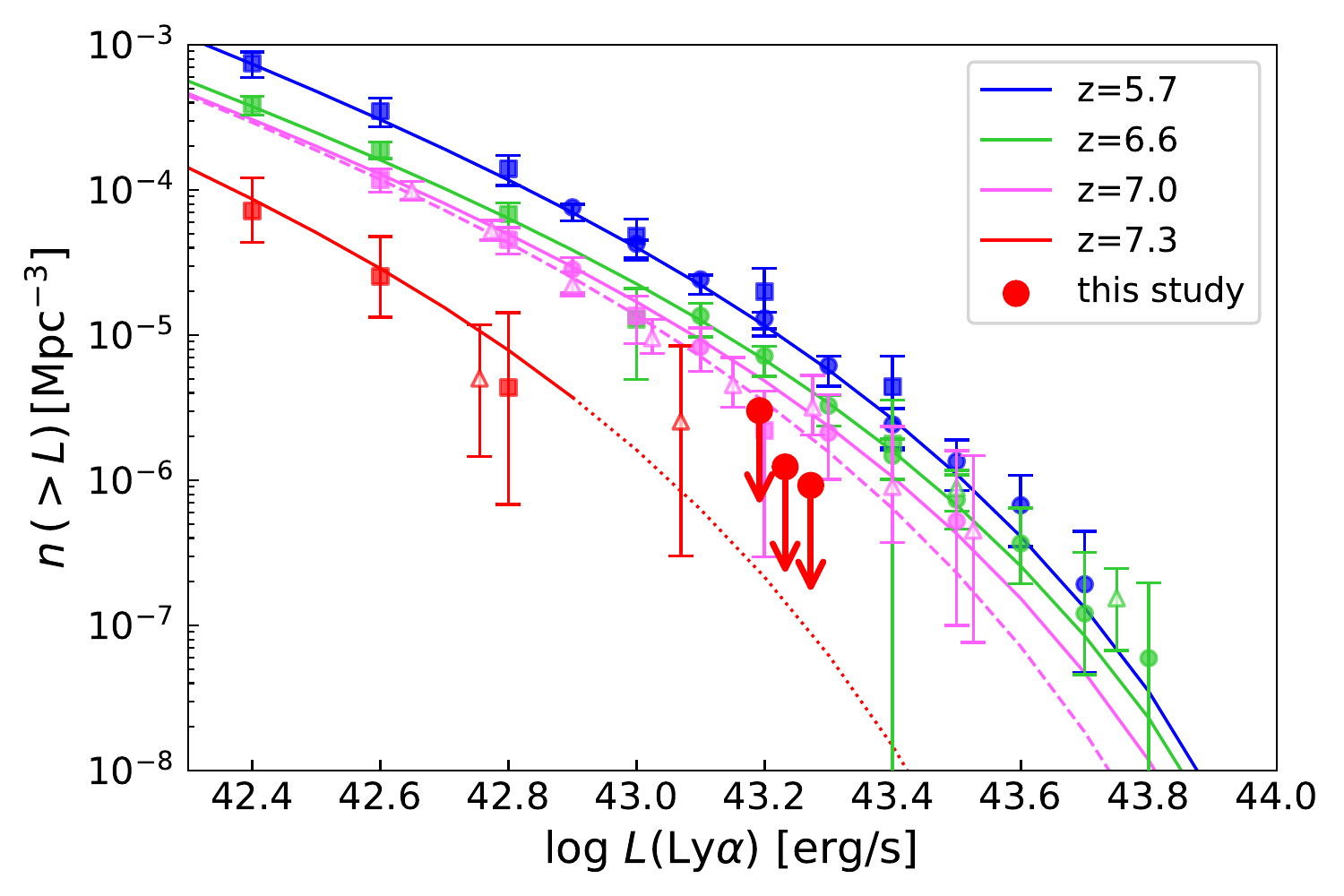}
    \caption{Cumulative Ly$\alpha$ LFs. The red circles are the upper limits at $z=7.3$ obtained by this study. The blue, green, and pink circles represent $z=5.7$, 6.6 \citep{Konno2018}, and 7.0 \citep{Itoh2018} Ly$\alpha$ LF measurements with HSC data. The blue, green, pink, and red squares represent $z=5.7$ \citep{Ouchi2008}, 6.6 \citep{Ouchi2010}, 7.0 \citep{Ota2017}, and 7.3 \citep{Konno2014} Ly$\alpha$ LF measurements with Subaru/Suprime-Cam data. The pink open triangles represent $z=7.0$ \citep{Hu2019} Ly$\alpha$ LF measurements. The green and red triangles represent $z=6.6$ \citep{Taylor2020} and $z=7.3$ \citep{Shibuya2012} Ly$\alpha$ LF measurements, respectively. The best-fit Schechter functions reported in these previous studies are shown by a blue solid line ($z=5.7$; \citealt{Konno2018}), a green solid line ($z=6.6$; \citealt{Konno2018}), a pink solid line ($z=7.0$; \citealt{Itoh2018}), a pink dashed line ($z=7.0$; \citealt{Hu2019}), and a red solid line ($z=7.3$; calculated by \citealt{Itoh2018} using the data given by \citealt{Konno2014} with a fixed faint-end slope of $\alpha=-2.5$; the bright part is shown by a dotted line because of no data).}
    \label{fig:cumLF}
\end{figure*}

In Figure \ref{fig:cumLF}, we show upper limits for three $L_{\Lya}$ values: $\log L_{\Lya}\ [\ergs] =43.19$ (23 {\tt patches}), 43.23 (56), and 43.27 (75).
The number in each parenthesis is the number of {\tt patches} used, which is dependent on $L_{\Lya}$ because different {\tt patches} have different limiting magnitudes (Figure \ref{fig:limitmag}).
We have searched for $L_{\Lya}$ (and the corresponding effective survey area) that gives the most stringent upper limit of $\TLya(7.3)/\TLya(5.7)$, finding that $\log L_{\Lya}\ [\ergs] =43.23$ (the data point in the middle) is the one. The results of the other two luminosities are plotted to show how much the upper limit changes with a slight change of $L_{\Lya}$.

To estimate Ly$\alpha$ limiting luminosities from the limiting magnitudes, we assume spectral energy distributions that have a flat ($f_\nu =$ const.) continuum, $\delta$-function Ly$\alpha$ emission with EW$_0 =100$ \AA, and zero flux at the wavelength bluer than Ly$\alpha$ due to the IGM absorption. This EW$_0$ value results in a conservative estimate of $L_{\Lya}$
(see, e.g., \citealt{Hashimoto2019} for the EW distribution of $z\sim6-8$ LAEs).

Figure \ref{fig:cumLF} shows the upper limits from this study together with the cumulative Ly$\alpha$ LFs of previous studies.
Our results are the first constraints on the bright ($\log L_{\Lya}\ [\ergs] \gtrsim 43$) part of the $z=7.3$ LF, making it possible to evaluate the IGM transmission using bright LAEs.
Our upper limits show a decrease from the Ly$\alpha$ LFs at $z=7.0$ derived by \cite{Itoh2018} (pink solid line and filled circles) and \cite{Hu2019} (pink dashed line and open triangles).

\subsection{IGM Transmission to Ly$\alpha$ photons}
\label{sec:T}
In this section, we derive the transmission of Ly$\alpha$ through the IGM, $\TLya(z)$, from the luminosity decrease of the Ly$\alpha$ LF.

The evolution of the Ly$\alpha$ LF is a combination of two effects: galaxy evolution (i.e., the intrinsic evolution of LAEs) and the change in $\TLya$ due to cosmic reionization.
To obtain implications for cosmic reionization, we need to resolve the degeneracy of these two effects.
\cite{Ouchi2010} have evaluated the effect of galaxy evolution using the UV LF evolution of LBGs.
The UV LF of LBGs also decreases from $z\sim 6$ to $z\sim 8$ (e.g., \citealt{Bouwens2015};  \citealt{Finkelstein2015}), suggesting that the cosmic star formation rate of galaxies declines over this redshift range.
In this study, we also estimate the effect of galaxy evolution with the same idea.

We assume that the observed $L_{\Lya}$ of galaxies can be written using their $L_\mathrm{UV}$ as:
\begin{eqnarray}
L_{\Lya} = \TLya(z) \fLya \kappa L_\mathrm{UV},
\label{eq:LLya-LUV}
\end{eqnarray}
where $\fLya$ is the Ly$\alpha$ escape fraction through the interstellar medium of galaxies and $\kappa$ is the Ly$\alpha$ production rate per UV luminosity.
The assumption that $\TLya$ is independent of intrinsic Ly$\alpha$ luminosity means that the observed Ly$\alpha$ luminosities of galaxies are uniformly decreased by IGM absorption, 
i.e., IGM absorption does not change the shape of the Ly$\alpha$ LF and only changes the characteristic luminosity, $L^*$.
We also assume that $\fLya$ and $\kappa$ do not change with redshift or UV luminosity, implying that the intrinsic Ly$\alpha$ LF of LAEs evolves in the same manner as the UV LF of LBGs.

\begin{figure*}[tb!]
    \centering
    \includegraphics[width=0.8\linewidth]{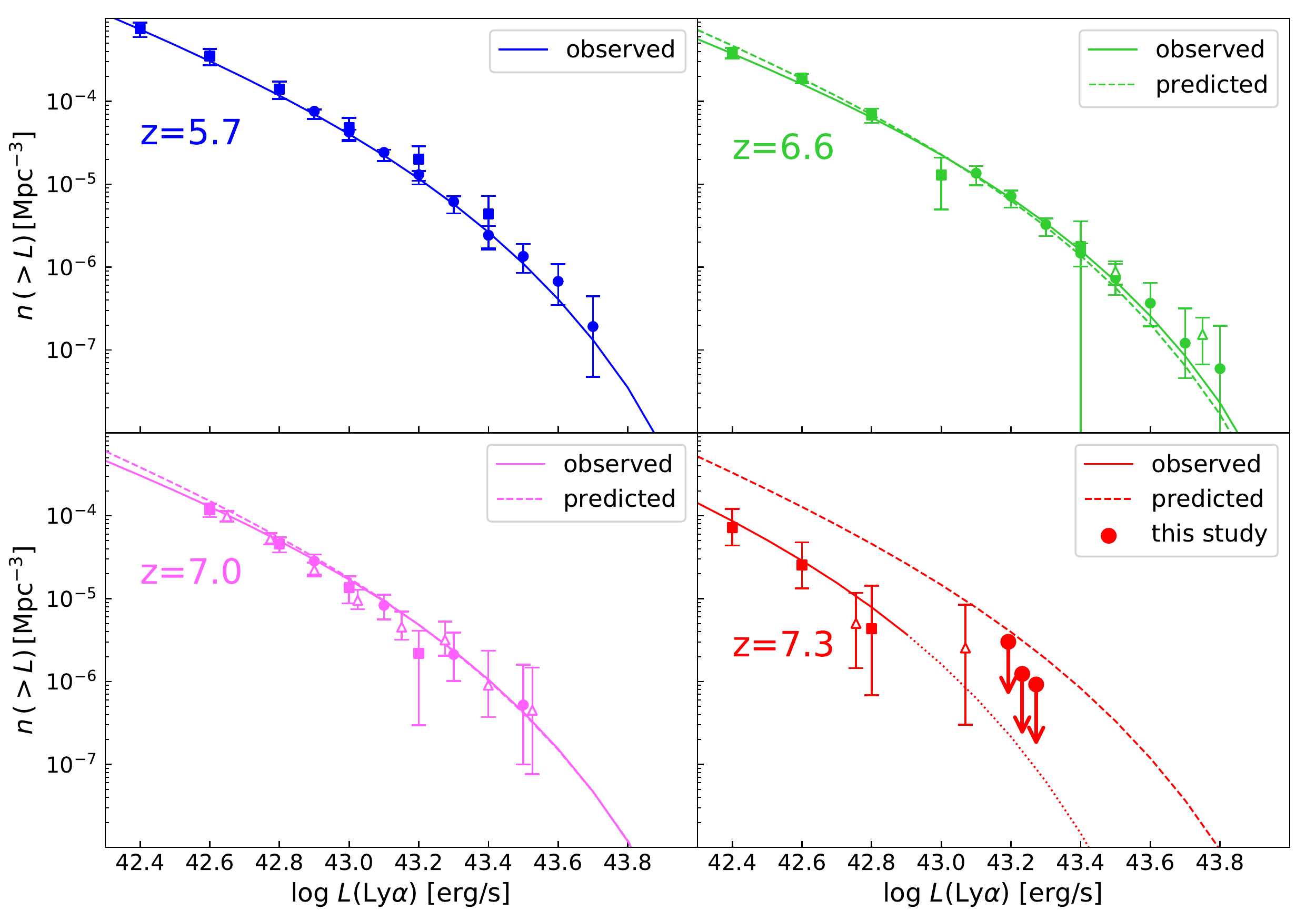}
    \caption{Comparison between the observed Ly$\alpha$ LFs (solid lines) and the predicted Ly$\alpha$ LFs (i.e., LFs for the fully ionized  IGM predicted by the evolution of the UV LF; see Section \ref{sec:T} for more details; dashed lines). 
    The meanings of symbols are the same as those in 
    Figure \ref{fig:cumLF}.}
    \label{fig:pred_obs}
\end{figure*}

To estimate $\TLya(z)/\TLya(5.7)$, we first predict the Ly$\alpha$ LF with the fully ionized IGM from the evolution of the UV LF.
A Schechter function \citep{Schechter1976} is defined by
\begin{eqnarray}
\phi(L)dL = \phi^* (L/L^*)^\alpha \exp(-L/L^*)d(L/L^*),
\end{eqnarray}
where $L^*$ is the characteristic luminosity, $\phi^*$ is the characteristic number density, and $\alpha$ is the faint-end slope.
We calculate the Schechter parameters of the predicted Ly$\alpha$ LF with the fully ionized IGM as:
\begin{eqnarray}
L^{* \mathrm{pred}}_{\Lya}(z) &=& L^{*  \mathrm{obs}}_{\Lya}(5.7) \times L^{*  \mathrm{obs}}_{\mathrm{UV}}(z) / L^{* \mathrm{obs}}_{\mathrm{UV}}(5.7), \nonumber\\
\phi^{* \mathrm{pred}}_{\Lya}(z) &=& \phi^{*  \mathrm{obs}}_{\Lya}(5.7) \times \phi^{*  \mathrm{obs}}_{\mathrm{UV}}(z) / \phi^{* \mathrm{obs}}_{\mathrm{UV}}(5.7), \nonumber\\
\alpha^\mathrm{pred}_{\Lya}(z) &=& \alpha^\mathrm{obs}_{\Lya}(5.7) + \alpha^\mathrm{obs}_{\mathrm{UV}}(z) - \alpha^\mathrm{obs}_{\mathrm{UV}}(5.7),
\label{eq:schechter}
\end{eqnarray}
where the superscripts {\lq}pred{\rq} and {\lq}obs{\rq} mean predicted and observed values, respectively.
This equation assumes that the UV LF evolves as described by the empirical model of \cite{Bouwens2015} (the first equation in their Section 5.1).
Specifically, we assume that $L^*_{\Lya}$ and $\phi^*_{\Lya}$ increase or decrease in the same ratio as those of the UV LF, and that $\alpha_{\Lya}$ increases or decreases additively in the same way as the UV LF.
For these calculations, we use the Schechter parameters of \cite{Konno2018} for the observed Ly$\alpha$ LF at $z=5.7$ and \cite{Bouwens2015} for the observed UV LFs.
Figure \ref{fig:pred_obs} shows a comparison between the predicted and observed Ly$\alpha$ LFs at $z=6.6$, 7.0, and 7.3.
We find that the observed Ly$\alpha$ LF follows the predicted one (and hence the UV LF) up to $z=7.0$ and then moves down at $z=7.3$.

\begin{figure}[htb!]
    \centering
    \includegraphics[width=\linewidth]{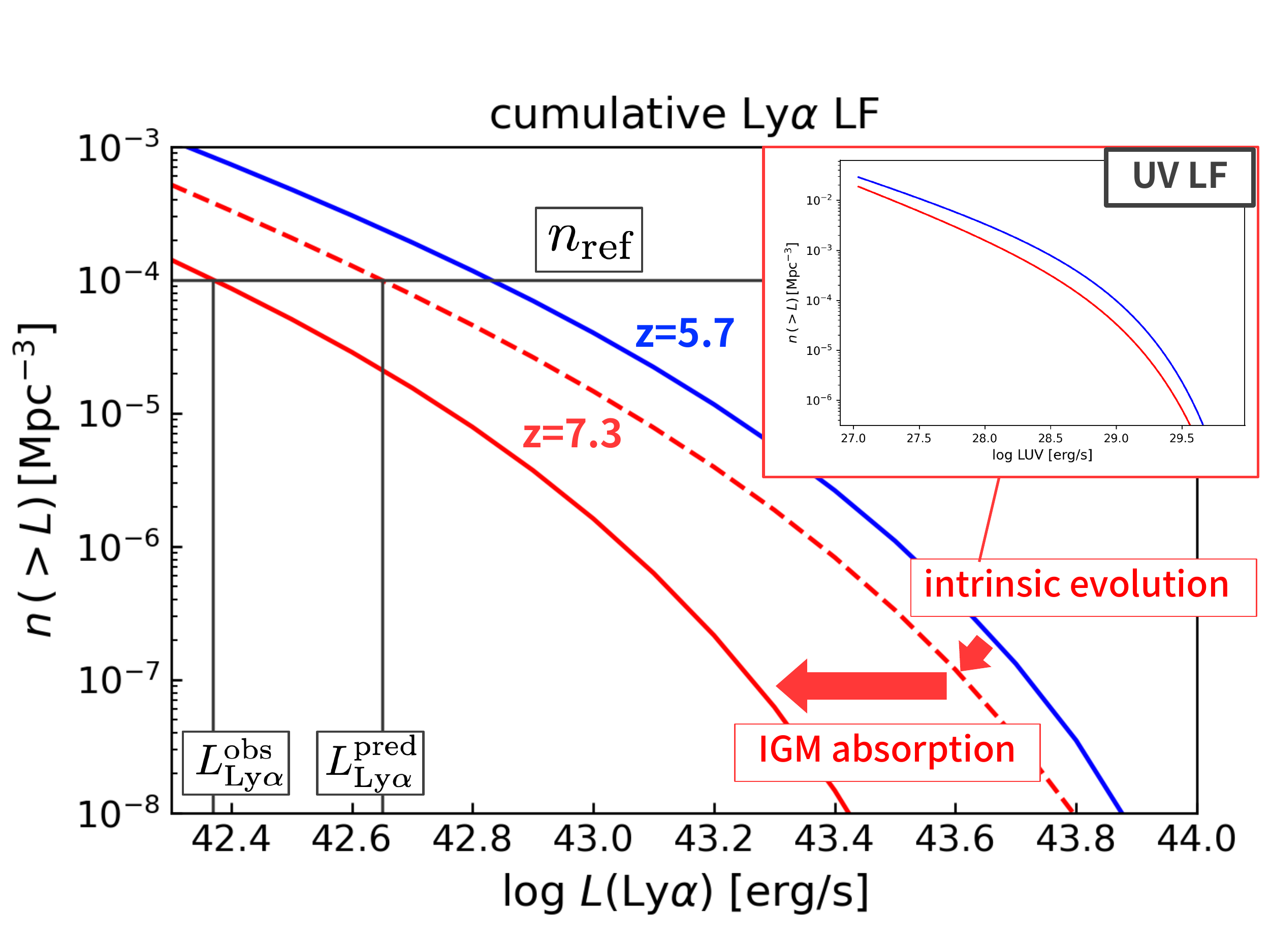}
    \caption{Schematic illustration of the method to measure the Ly$\alpha$ transmission of the IGM, $\TLya$. The blue and red solid lines indicate the observed Ly$\alpha$ LFs at $z=5.7$ and 7.3, respectively. The red dashed line is the intrinsic Ly$\alpha$ LF at $z=7.3$, which is predicted using the observed $z=5.7$ Ly$\alpha$ LF on the assumption that the intrinsic Ly$\alpha$ LF evolves in the same way as the UV LF (inset figure).}
    \label{fig:T_calc}
\end{figure}

We then calculate $\TLya(z)/\TLya(5.7)$ by measuring the luminosity decrease between the predicted and observed Ly$\alpha$ LFs.
Previous studies have used the luminosity density to evaluate $\TLya$ \citep[e.g.,][]{Konno2018, Itoh2018, Hu2019}, but this method systematically underestimates the luminosity decrease because of a fixed integration range of the LF, as explained in Appendix \ref{appendix:B}.
Therefore, we directly measure the luminosity decrease using another method (Figure \ref{fig:T_calc}).
First, we set a reference cumulative number density, $n_\mathrm{ref}$.
Then, we look for the Ly$\alpha$ luminosities ($L^\mathrm{obs}_{\Lya}(z)$ for the observed Ly$\alpha$ LF and $L^\mathrm{pred}_{\Lya}(z)$ for the predicted one) that satisfy
$\int_{L^\mathrm{obs}_{\Lya}}^\infty \phi^\mathrm{obs}(L)dL = \int_{L^\mathrm{pred}_{\Lya}}^\infty \phi^\mathrm{pred}(L)dL = n_\mathrm{ref}$.
We calculate $\TLya(z)/\TLya(5.7)$ as:
\begin{eqnarray}
\frac{\TLya(z)}{\TLya(5.7)} = \frac{L^\mathrm{obs}_{\Lya}(z)}{L^\mathrm{pred}_{\Lya}(z)}.
\label{eq:T_L}
\end{eqnarray}
In this way, we calculate $\TLya(z)/\TLya(5.7)$ from the most stringent upper limit of this study at $z=7.3$.
We also apply the same calculation to the Ly$\alpha$ LFs derived by previous studies at $z=6.6$ \citep{Konno2018}, 7.0 (\citealt{Itoh2018}; \citealt{Hu2019}), and 7.3 \citep{Konno2014}.\footnote{In our analysis, we use literature Ly$\alpha$ LF measurements that have derived the best-fit Schechter parameters.
Our analysis does not include \cite{Shibuya2012} and \cite{Taylor2020} because of their limited data points of the Ly$\alpha$ LF. Their data points are consistent with the Ly$\alpha$ LFs at similar redshifts used in this study (Figure \ref{fig:cumLF}).
We do not use \cite{Santos2016}, either. \cite{Santos2016} have reported a higher number density than \cite{Konno2018} at $z=5.7$ and $6.6$. The reason for this discrepancy is unclear, but one possible explanation is that their completeness correction is redundant \citep{Konno2018}. In this study, we adopt \cite{Konno2018} whose completeness correction method is the same as ours.}
At each redshift, we calculate $\TLya(z)/\TLya(5.7)$ from a bright part and a faint part of the Ly$\alpha$ LF, to examine if different parts of the LF give consistent $\TLya(z)/\TLya(5.7)$ values. If not, it implies either that the actual Ly$\alpha$ LF does not obey Equation (\ref{eq:schechter}) or that $\TLya$ is not independent of Ly$\alpha$ luminosity.
We adopt $n_\mathrm{ref} = 1 \times 10^{-6}\ \mathrm{Mpc}^{-3}$ for a bright part because, at this value, our $z=7.3$ data can place the most stringent upper limit on the IGM transmission. 
For a faint part, we adopt $n_\mathrm{ref} = 1 \times 10^{-4}\ \mathrm{Mpc}^{-3}$, the highest value where LF measurements are available for all four redshifts.
These $n_\mathrm{ref}$ values are common to $z=6.6$, 7.0, and 7.3.

\begin{figure}[tb!]
    \centering
    \includegraphics[width=\linewidth]{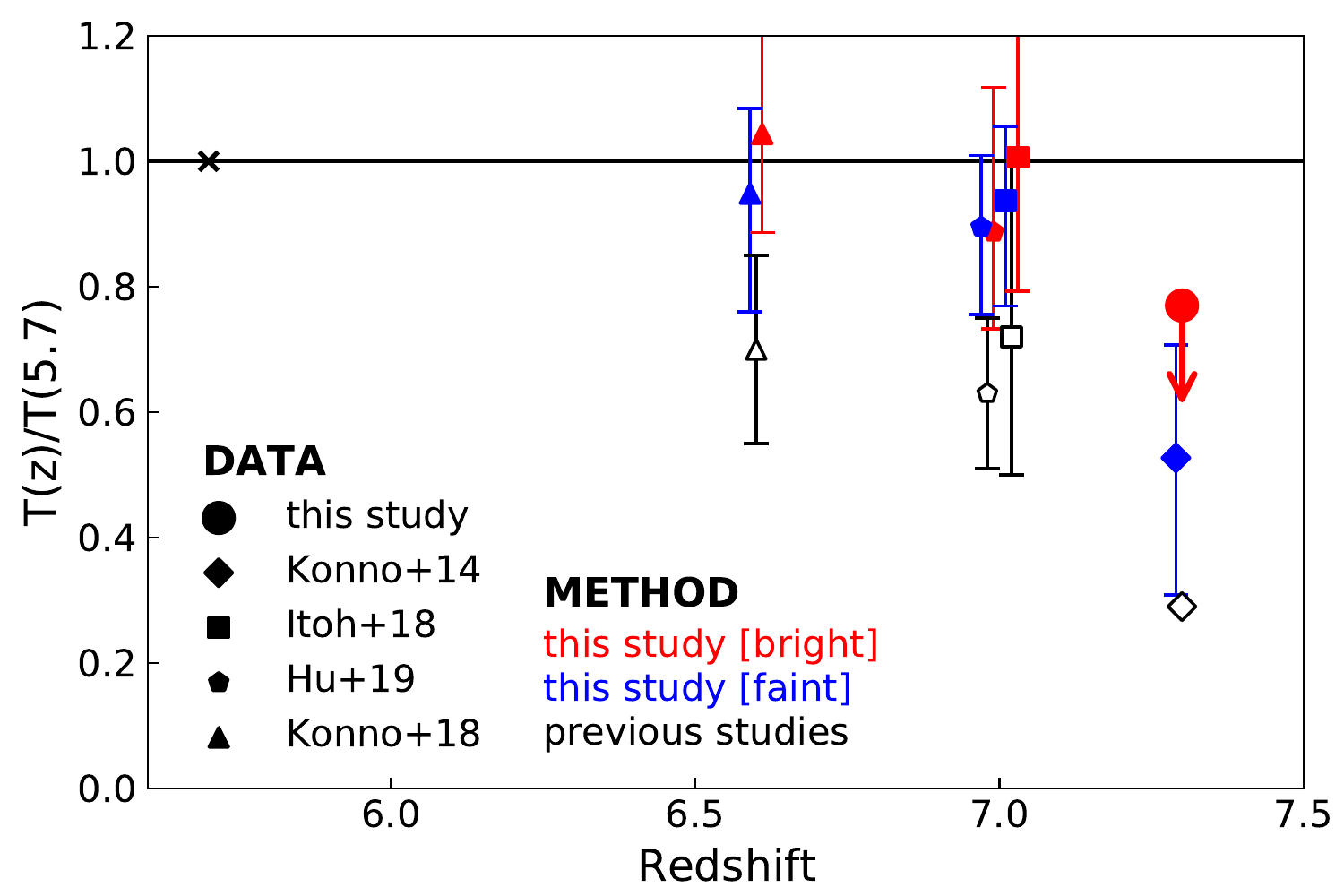}
    \caption{Ly$\alpha$ Transmission, $\TLya(z)/\TLya(5.7)$, as a function of redshift. 
    The red and blue symbols indicate, respectively, the results for bright and faint parts of the LF at each redshift calculated by this study's new method:
    a circle at $z=7.3$, our new data; 
    a diamond at $z=7.3$, \cite{Konno2014};
    squares at $z=7.0$, \cite{Itoh2018}; 
    pentagons at $z=7.0$, \cite{Hu2019}; 
    and triangles at $z=6.6$, \cite{Konno2018}.
    Also plotted in black symbols are the values obtained in the previous studies using luminosity densities. The horizontal line represents $\TLya(z)/\TLya(5.7)=1$. Points are slightly offset in the redshift direction for clarity.}
    \label{fig:T}
\end{figure}

Figure \ref{fig:T} and Table \ref{tab:T,xHI} show the values of $\TLya(z)/\TLya(5.7)$ thus obtained.
Error bars include uncertainties from the Ly$\alpha$ LFs at that redshift and $z=5.7$ and the UV LF evolution.
To estimate the uncertainties from the UV LF evolution, we use the first equation in Section 5.1 of \cite{Bouwens2015}.
From the upper limit of this study, we obtain $\TLya(7.3)/\TLya(5.7) < 0.77$, and from the Ly$\alpha$ LFs of previous studies, we obtain $\TLya(6.6, 7.0)/\TLya(5.7) \simeq 1$ and $\TLya(7.3)/\TLya(5.7) = 0.53^{+0.18}_{-0.22}$.
The bright and faint parts give almost the same results at $z=6.6$ and 7.0,\footnote{For \cite{Konno2014}, which have no data in the bright part, we calculate the luminosity decrease only in the faint part.}
which is consistent with our assumption that the intrinsic Ly$\alpha$ LF of LAEs evolves in the same way as the UV LF of LBGs and that the effect of IGM absorption, $\TLya$, does not depend on Ly$\alpha$ luminosity (Equation (\ref{eq:LLya-LUV})).
We also plot $\TLya(z)/\TLya(5.7)$ calculated in previous studies using luminosity densities.
The measurements of \cite{Konno2018}, \cite{Itoh2018}, and \cite{Hu2019}, which adopted an integration range of $\log L_{\Lya}\ [\ergs] =$ 42.4--44, are lower than our results as expected (see Appendix \ref{appendix:B}).

\begin{deluxetable*}{clccccc}
\tablecaption{Summary of $\TLya$ and $\xHI$ estimates\label{tab:T,xHI}}
\tablehead{\colhead{$z$} & \colhead{Ly$\alpha$ LF} & \colhead{$L^*$} & \colhead{$\phi^*$} & \colhead{$\alpha$} & \colhead{$\TLya(z)/\TLya(5.7)$} & \colhead{$\xHI$}\\
&& \colhead{($10^{43}\ \ergs$)} & \colhead{($10^{-4}\  \mathrm{Mpc}^{-3}$)} &&& \\
\colhead{(1)} & \colhead{(2)} & \colhead{(3)} & \colhead{(4)} &  \colhead{(5)} & \colhead{(6)} & \colhead{(7)}}
\startdata
7.3 & this study & --- & --- & --- & $<0.77$ & $>0.28$\\
& \cite{Konno2014}$^\ddag$ & $0.55^{+9.45}_{-0.33}$ & $0.94^{+12.03}_{-0.93}$ & $-2.5$ (fixed)& $0.53^{+0.18}_{-0.22}$$^\dag$ & $0.39^{+0.12}_{-0.08}$$^\dag$\\
7.0 & \cite{Itoh2018} & $1.50^{+0.42}_{-0.31}$ & $0.45^{+0.26}_{-0.18}$ & $-2.5$ (fixed) & $0.94^{+0.12}_{-0.17}$$^\dag$ & $0.16^{+0.14}_{-0.16}$$^\dag$\\
& \cite{Hu2019} & $1.20^{+0.46}_{-0.27}$ & $0.65^{+0.52}_{-0.33}$ & $-2.5$ (fixed) & $0.90^{+0.11}_{-0.14}$$^\dag$ & $0.20^{+0.11}_{-0.20}$$^\dag$\\
6.6 & \cite{Konno2018} & $1.66^{+0.30}_{-0.69}$ & $0.467^{+1.44}_{-0.442}$ & $-2.49^{+0.50}_{-0.50}$ & $0.95^{+0.14}_{-0.19}$$^\dag$ & $0.15^{+0.16}_{-0.15}$$^\dag$\\
5.7 & \cite{Konno2018} & $1.64^{+2.16}_{-0.62}$ & $0.849^{+1.87}_{-0.771}$ & $-2.56^{+0.53}_{-0.45}$ & --- & ---
\enddata
\tablecomments{
(1) Redshift.
(2) Ly$\alpha$ LF used to calculate $\TLya(z)/\TLya(5.7)$.
(3) Characteristic luminosity of the Ly$\alpha$ LF.
(4) Characteristic number density of the Ly$\alpha$ LF.
(5) Faint-end slope of the Ly$\alpha$ LF.
(6) Transmission of Ly$\alpha$ through the IGM obtained in Section \ref{sec:T}.
(7) Volume-averaged neutral hydrogen fraction in the IGM obtained in Section \ref{sec:xHI}.}
\tablenotetext{\dag}{Values calculated from the faint part of the Ly$\alpha$ LF ($n_\mathrm{ref} = 1 \times 10^{-4}\  \mathrm{Mpc}^{-3}$; see Section \ref{sec:T}).}
\tablenotetext{\ddag}{We use the Ly$\alpha$ LF calculated by \cite{Itoh2018} using the data given by \cite{Konno2014}.}
\end{deluxetable*}

\subsection{IGM Neutral Hydrogen Fraction}
\subsubsection{Estimation of $\xHI$}
\label{sec:xHI}
From the  $\TLya(z)/\TLya(5.7)$ obtained in Section \ref{sec:T}, we estimate the volume-averaged neutral hydrogen fraction in the IGM, $\xHI$\footnote{Hereafter, we refer to the volume-averaged neutral hydrogen fraction as $\xHI$.}, in the same manner as \cite{Jung2020} assuming inhomogeneous reionization.
Theoretically, $\TLya(z)$ is described as:
\begin{eqnarray}
&\TLya(z) = e^{-\tau_\mathrm{IGM}},\nonumber\\
&\tau_\mathrm{IGM} = \tau_\mathrm{D} + \tau_\mathrm{HII},
\label{eq:tauIGM}
\end{eqnarray}
where $\tau_\mathrm{IGM}$ is the total optical depth of the IGM, $\tau_\mathrm{D}$ is the optical depth of neutral patches, and $\tau_\mathrm{HII}$ is the optical depth of ionized bubbles \citep{Dijkstra2014}.
We assume that $\tau_\mathrm{HII}$ does not change with redshift, which leads to
\begin{eqnarray}
\frac{\TLya(z)}{\TLya(5.7)} = e^{-\tau_\mathrm{D}(z)},
\label{eq:T-tauD}
\end{eqnarray}
assuming $\tau_\mathrm{D}(5.7)=0$.

To obtain $\xHI$, we use an analytical approach of \cite{Dijkstra2014} that considers inhomogeneous reionization (his Equation (30)):
\begin{eqnarray}
\tau_D (z_g, \Delta v) &\approx& 2.3 \xHI \left( \frac{\Delta v_b}{600\ \mathrm{km\ s^{-1}}}\right)^{-1} \left(\frac{1+z_g}{10}\right)^{3/2}, \nonumber \\
\Delta v_b &=& \Delta v + H(z_g) R_b / (1+z_g),
\label{eq:tau-Rb}
\end{eqnarray}
where $z_g$ is the systemic redshift of a galaxy, $\Delta v$ is the velocity offset of the galaxy's Ly$\alpha$ emission from the systemic redshift, $\Delta v_b$ is the velocity offset from line resonance when the Ly$\alpha$ photons from the galaxy first enter a neutral patch, $H(z_g)$ is the Hubble constant at $z_g$, and $R_b$ is the comoving distance to the surface of the neutral patch.
If we adopt for $\Delta v$ the typical range for $z\sim 6-8$ LAEs obtained by \cite{Hashimoto2019},  
$\Delta v=100^{+100}_{-100} \ \mathrm{km\ s^{-1}}$, then the unknown quantities are $\xHI$ and $R_b$.

We then use the characteristic size of ionized bubbles, $R_b$, as a function of $z$ predicted by \cite{Furlanetto2005} with an analytic model of patchy reionization;
we calculate the $\xHI$ -- $R_b$ relation at $z_g =6.6, 7.0$, and $7.3$ by interpolating the relations at $z=6$ and 9 in Figure 1 of \cite{Furlanetto2005}.

\begin{figure}[tb!]
    \centering
    \includegraphics[width=\linewidth]{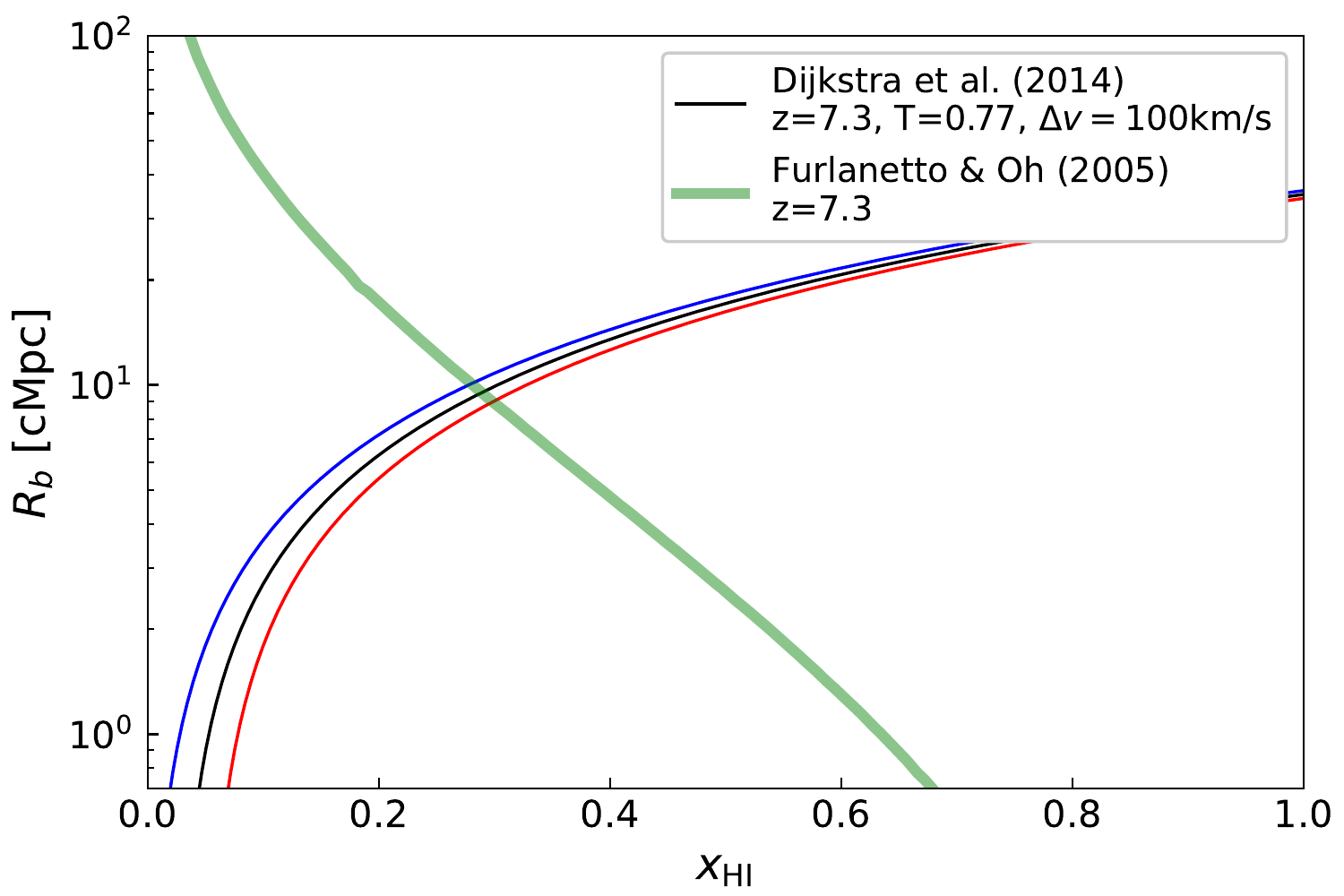}
    \caption{
    Estimation of the volume-averaged neutral hydrogen fraction in the IGM ($\xHI$) using our NB1010 data (Section \ref{sec:xHI}).
    The blue, black, and red lines represent the calculations from Equations (\ref{eq:T-tauD}) and (\ref{eq:tau-Rb}) (i.e., from an analytical approach of \citealt{Dijkstra2014}) with $\Delta v=0$, 100, and 200 $\mathrm{km\ s^{-1}}$, respectively, for $\TLya(7.3)/\TLya(5.7)=0.77$.
    The green thick line represents the characteristic size of ionized bubbles ($R_b$) predicted by \cite{Furlanetto2005}.}
    \label{fig:xHI-R}
\end{figure}

Figure \ref{fig:xHI-R} shows the $\xHI$ -- $R_b$ relation from \cite{Dijkstra2014} for $\TLya(7.3)=0.77$ as an example.
The blue, black, and red lines represent the calculation from Equations (\ref{eq:T-tauD}) and (\ref{eq:tau-Rb}) (i.e., \citealt{Dijkstra2014}) with $\Delta v=0$, 100, and 200 $\mathrm{km\ s^{-1}}$, respectively, which indicates that larger $\Delta v$ and $R_b$ result in higher $\xHI$ because of an easier escape of Ly$\alpha$ photons.
On the other hand, the green line is the prediction by \cite{Furlanetto2005}, which shows a larger $R_b$ in a more ionized (lower $\xHI$) universe.
We obtain $\xHI$ as the intersection of these two lines and conservatively evaluate its uncertainty following \cite{Jung2020}, allowing a range of $\Delta v=0$ -- $200 \ \mathrm{km\ s^{-1}}$ (see Figure 12 of \citealt{Jung2020}).
In this way, we calculate $\xHI(z)$ for all the  $\TLya(z)/\TLya(5.7)$ measurements obtained in Section \ref{sec:T}.

From our new data, we obtain $\xHI>0.28$ at $z=7.3$. From the literature Ly$\alpha$ LFs, we obtain $\xHI=0.39_{-0.08}^{+0.12}$ at $z=7.3$, and $\xHI$ consistent with zero within the errors at $z=6.6$ and 7.0 (Figure \ref{fig:xHI} and Table \ref{tab:T,xHI}).
Since these $\xHI$ estimates are based on specific models of Ly$\alpha$ transmission in the IGM and the evolution of ionized bubbles, we also estimate $\xHI$ using two other models and obtain consistent results (Appendix \ref{appendix:C}).

\subsubsection{Possible uncertainties in the $\xHI$ estimates}
\label{sec:uncertainties}
In this section, we discuss possible uncertainties in our $\xHI$ estimates.
First, if $\kappa$ and/or $\fLya$ in Equation (\ref{eq:LLya-LUV}), which we assume to be constant in Section \ref{sec:T}, increases at $z>5.7$, the acutual $\xHI$ would be larger than our results.
If $\kappa$ or $\fLya$ increases, the emitted Ly$\alpha$ luminosity also increases; thus, the luminosity decrease due to $\xHI$ needs to be greater to reproduce the observed Ly$\alpha$ LF, which results in higher $\xHI$.
Indeed, \cite{Hayes2011} have found that $\fLya$ increases with redshift over $0<z<6$. 
It is, however, not clear whether $\fLya$ and $\kappa$ significantly increase 
from $z=5.7$ to $z=7.0$ (or to $z=7.3$), a period shorter than 300 Myr.
We also note that the Ly$\alpha$ EW method also adopts essentially the same assumption.

Second, bright LAEs targeted in this study may be in larger ionized bubbles than the average ones adopted in Section \ref{sec:xHI}, 
implying that we may be underestimating $\xHI$.
Figure 2 of \cite{Furlanetto2005} shows bubble size distributions for different total masses, 
with the most massive regions having three times larger sizes than the average.
Using the three times larger size in Section \ref{sec:xHI} will give $\xHI(7.3) \gtrsim 0.4$.

Finally, previous simulations \citep{Mesinger2008, Dijkstra2011, Mason2018b, Weinberger2019} of LAEs during reionization demonstrate that $\TLya$ has a broad distribution at a given $\xHI$ due to a broad range of ionized bubble sizes and a sightline-to-sightline scatter.
This effect gives an additional uncertainty to $\xHI$ estimates using $\TLya$.
However, applying this effect to Ly$\alpha$ LF-based $\xHI$ estimates is complicated and beyond the scope of this paper.

We also note that because our analysis uses the integrated luminosity density, any information on the shape of the Ly$\alpha$ LF is lost.
An accurate determination of the LF shape from a deeper and larger LAE survey 
may place some constraints on the topology of reionization 
through, e.g., the dependence of bubble sizes on Ly$\alpha$ luminosity.

\subsubsection{Comparison with previous studies}
\label{sec:comparison}
\begin{figure*}[tb!]
    \centering
    \includegraphics[width=0.8\linewidth]{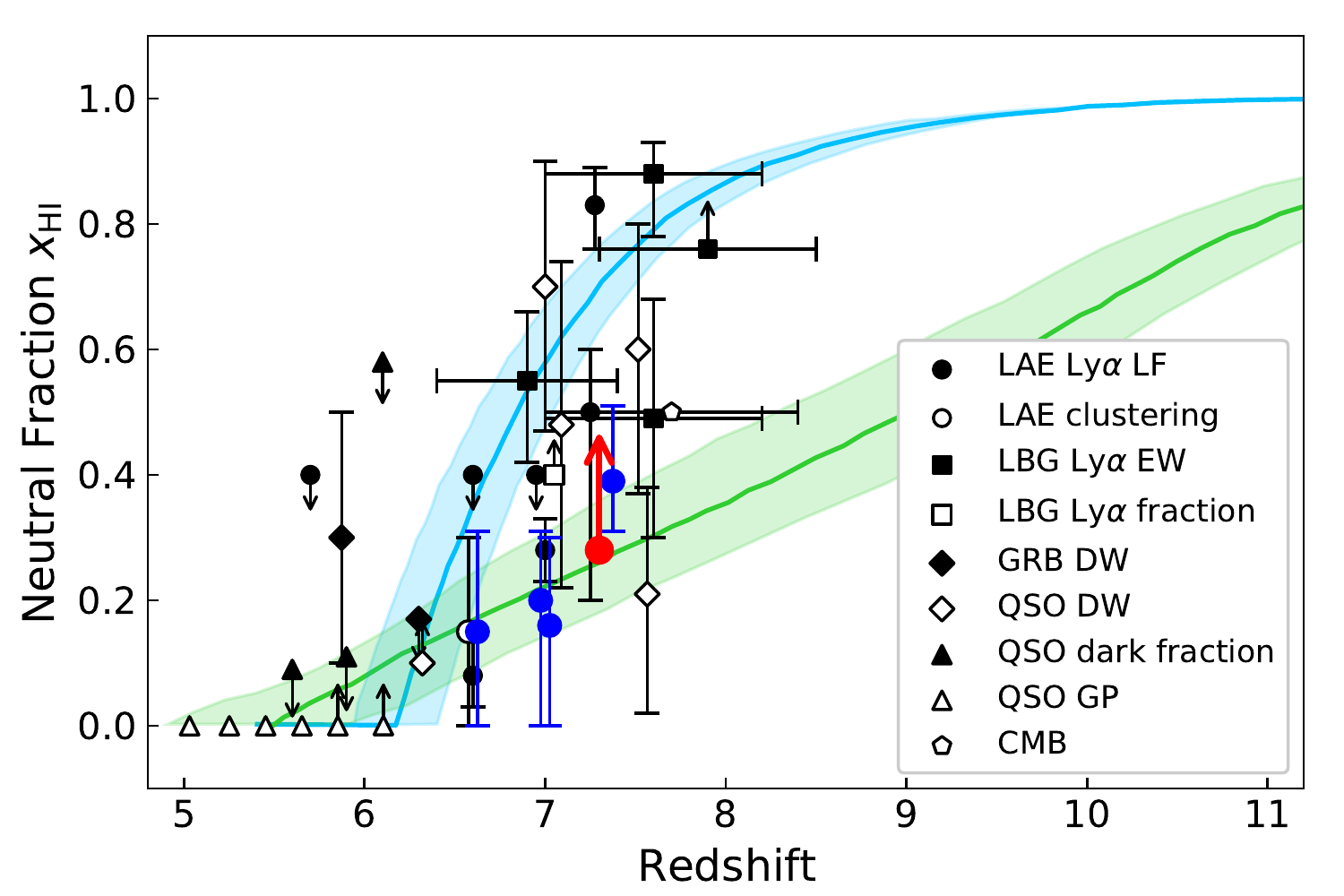}
    \caption{Redshift evolution of the volume-averaged neutral hydrogen fraction in the IGM, $\xHI$. Our new constraints based on the Ly$\alpha$ LF are shown by a red filled circle (estimated from our new data) and blue filled circles (estimated from the previous studies' Ly$\alpha$ LFs). We also plot estimates derived from the Ly$\alpha$ LF \citep[filled circle;][]{Inoue2018, Morales2021}; the clustering of LAEs \citep[open circle;][]{Ouchi2018}; the Ly$\alpha$ EW distribution of LBGs \citep[filled squares;][]{Hoag2019, Mason2019, Whitler2020, Jung2020}; the fraction of LBGs emitting Ly$\alpha$ \citep[Ly$\alpha$ fraction; open square;][]{Mesinger2015}; GRB damping wings \citep[filled diamonds;][]{Totani2006, Totani2014}; QSO damping wings \citep[open diamonds;][]{Schroeder2013, Davies2018, Greig2019, Wang2020}; Ly$\alpha$ and Ly$\beta$ forest dark fractions of QSOs \citep[filled triangles;][]{McGreer2015}; the Gunn-Peterson trough of QSOs \citep[open triangles;][]{Fan2006}; and the CMB Thomson optical depth \citep[open pentagon;][]{Planck2020}. We slightly offset the constraints at $z=6.6$, 7.0, 7.3, and 7.54 in the redshift direction for clarity. Two semi-empirical reionization models are also plotted in a green line \citep{Finkelstein2019} and a light blue line (Model I of \citealt{Naidu2020}).}
    \label{fig:xHI}
\end{figure*}

Figure \ref{fig:xHI} and Table \ref{tab:T,xHI} show the estimates of $\xHI$ from our new data (red symbol in Figure \ref{fig:xHI}) and the previous studies' Ly$\alpha$ LFs (blue symbols).
Also plotted in Figure \ref{fig:xHI} are other estimates in the literature (black symbols). 

First, we focus on the Ly$\alpha$ LF-based $\xHI$ estimates.
At $z=7.3$, we obtain $\xHI>0.28$ from our new data.
This lower limit is consistent with our estimate from \cite{Konno2014}'s LF (the faint part of the $z=7.3$ Ly$\alpha$ LF),
and indicates that cosmic reionization is ongoing at $z\sim 7.3$.
On the other hand, 
the $\xHI(6.6)$ and $\xHI(7.0)$ values obtained from both the bright and faint parts of the corresponding LFs
are consistent with full ionization within the errors, indicating that the universe is completing reionization around these redshifts.
The estimates from this study are also consistent with those by \cite{Inoue2018}, $\xHI(7.3)=0.5_{-0.3}^{+0.1}$ and $\xHI(5.7, 6.6, 7.0)<0.4$.
They have predicted Ly$\alpha$ LFs in the fully ionized IGM not from observed UV LFs but by a physically motivated analytic model of LAEs that calculates Ly$\alpha$ luminosity as a function of dark halo mass.
Their model reproduces
observed Ly$\alpha$ LFs, LAE angular auto-correlation functions, and LAE fractions in LBGs at $z\sim 6-7$.
Very recently, \cite{Morales2021} have predicted Ly$\alpha$ LFs for various $\xHI$ in a partially ionized universe with an analytic model of the UV LF and infer the IGM neutral fraction at $z=6.6$, 7.0, and 7.3 from a comparison with observed Ly$\alpha$ LFs.
Their $\xHI(6.6)$ and $\xHI(7.0)$ are consistent with our results within the errors, but their $\xHI(7.3)$ is higher than ours from \cite{Konno2014}'s LF.
The cause of the difference at $z=7.3$ has not been fully identified, but it is partly because of the conversion from the decrease in the Ly$\alpha$ LF to $\xHI$.

Next, we compare these Ly$\alpha$ LF-based results with the other methods' results.
At $z\sim 7.0$, our constraints,  $\xHI(7.0)=0.16^{+0.14}_{-0.16}$ and $0.20^{+0.11}_{-0.20}$ from the Ly$\alpha$ LFs of \cite{Itoh2018} and \cite{Hu2019}, respectively, are lower than the other results,  $\xHI(7.0)=0.70^{+0.20}_{-0.23}$ (\citealt{Wang2020}; QSO damping wing), $\xHI(\sim 7)=0.55^{+0.11}_{-0.13}$ (\citealt{Mason2018a}; \citealt{Whitler2020}; LBG EW distribution), and $\xHI(\sim 7)>0.4$ (\citealt{Mesinger2015}; LBG Ly$\alpha$ fraction) despite a large uncertainty in each estimate.
If the Ly$\alpha$ LF-based estimates are underestimating $\xHI$, the cause could be the assumption of constant $\kappa$ and $\fLya$ and/or the use of the average size of ionized bubbles as mentioned in Section \ref{sec:uncertainties}.

Our constraint of $\xHI(7.3)>0.28$ is broadly consistent with the other estimates at $z\sim 7.0-7.6$
that span $0.2\lesssim \xHI \lesssim 0.9$, 
thus adding further evidence of reionization being still underway around $z=7.3$.
The estimate by \cite{Greig2019},
$\xHI(7.54)=0.21^{+0.17}_{-0.19}$,
is the lowest among the all estimates including ours (although within the errors).
Their source, QSO J1342, is the same as of \cite{Banados2018} and \cite{Davies2018}, who have obtained $\xHI\sim 0.6$,
but \cite{Greig2019} have analyzed only the red side of the observed Ly$\alpha$ spectrum to avoid complicated modelling of the near-zone transmission.
Therefore, different analyses can lead to largely different results even for the same source.
If the result of \cite{Greig2019} is correct, it is possible that this QSO resides in a large HII region.
Indeed, it has been suggested that QSOs inhabit highly biased overdense regions which were reionized early \citep[e.g.,][]{Mesinger2010, Dijkstra2014}.
A similarly large difference among the four Ly$\alpha$ EW-based estimates over $7\lesssim z \lesssim8$ may also be partly attributed to the presence or not of a highly ionized region as suggested by \cite{Jung2020}, although these estimates might be detecting a real change in $\xHI$ with a coarse resolution of $\Delta z\sim 1$.

In Figure \ref{fig:xHI}, we also plot semi-empirical models of reionization by \cite{Finkelstein2019} and \cite{Naidu2020}.
\cite{Finkelstein2019} have predicted early and smooth reionization driven by faint galaxies, with a steep faint-end slope ($\alpha_\mathrm{UV}<-2$) of the UV LF and higher escape fractions of ionizing photons ($\fion$) in fainter galaxies.
On the other hand, \cite{Naidu2020} have predicted late and rapid reionization driven by bright galaxies, with a shallow faint-end slope ($\alpha_\mathrm{UV}>-2$). 
For $f_\mathrm{esc}^\mathrm{ion}$, \cite{Naidu2020} have examined two cases:
one assuming a constant $f_\mathrm{esc}^\mathrm{ion}$ across all galaxies (Model I) and the other assuming $f_\mathrm{esc}^\mathrm{ion}$ 
to be dependent on the SFR surface density,  $\Sigma_\mathrm{SFR}$, of galaxies (Model II).
The difference in $\xHI$ between these two cases is relatively small.
Our $\xHI$ estimates seem to prefer \cite{Finkelstein2019}'s model to \cite{Naidu2020}'s.

In summary, we provide a new constraint of $\xHI(7.3)>0.28$ from NB-selected LAEs' Ly$\alpha$ LF.
This is a constraint from a large ($\sim 2 \times 10^6\ \mathrm{Mpc}^3$) cosmic volume with a negligibly small redshift uncertainty.
If the possible underestimation of the Ly$\alpha$ LF method
due, for example, to an increase in $\kappa$ or $\fLya$ is true, then the actual $\xHI$ will be even higher.

\section{Conclusions}
We have derived a new constraint on the Ly$\alpha$ LF of $z=7.3$ LAEs based on a large-area narrow-band imaging survey with Subaru/Hyper Suprime-Cam whose effective survey volume is $\sim 2\times 10^6$ Mpc$^3$. Using this constraint, we have calculated the Ly$\alpha$ transmission in the IGM, $\TLya$, and then the volume-averaged neutral hydrogen fraction in the IGM, $\xHI$, at $z=7.3$.
In the calculation of $\TLya$, we have applied a new method that directly measures the luminosity decrease between an observed LF and a predicted LF (i.e., LF for the fully ionized IGM predicted by the evolution of the UV LF).
We have also applied this method to previous studies' Ly$\alpha$ LFs at $z=6.6$, 7.0, and 7.3.
Our main results are summarized below.
\begin{enumerate}
    \item We have detected no $z=7.3$ LAEs in either the COSMOS or SXDS field, which results in a decrease in the bright part of the Ly$\alpha$ LF from $z=7.0$ \citep{Itoh2018, Hu2019} to $z=7.3$ (Figure \ref{fig:cumLF}).
    \item To estimate $\xHI$, we have predicted the Ly$\alpha$ LF in the case of the fully ionized IGM on the assumption that the intrinsic Ly$\alpha$ LF evolves in the same way as the observed UV LF.
    We have found that the observed Ly$\alpha$ LF follows the predicted one (and hence the UV LF) up to $z=7.0$ and then moves down at $z=7.3$ (Figure \ref{fig:pred_obs}).
    \item We have estimated $\TLya$ in a new method that directly measures the luminosity decrease between an observed Ly$\alpha$ LF and a predicted one.
    We have obtained $\TLya(7.3)/\TLya(5.7) < 0.77$ from our new data, and $\TLya(6.6, 7.0)/\TLya(5.7) \simeq 1$ and $\TLya(7.3)/\TLya(5.7) = 0.53^{+0.18}_{-0.22}$ from the previous studies' Ly$\alpha$ LFs (\citealt{Konno2014, Konno2018, Itoh2018, Hu2019}; Figure \ref{fig:T} and Table \ref{tab:T,xHI}).
    Bright and faint parts of the Ly$\alpha$ LF give almost the same $\TLya$.
    \item Using the obtained $\TLya$, we have estimated $\xHI$ in the same manner as \cite{Jung2020}.
    The constraint of $\xHI(7.3)>0.28$ estimated from our new data is broadly consistent with the other estimates in the literature and indicates that cosmic reionization is still ongoing at $z\sim7.3$.
    On the other hand, the $\xHI(6.6)$ and $\xHI(7.0)$ estimated from the previous studies' Ly$\alpha$ LFs are consistent with full ionization but are lower than the other estimates at the same redshift (\citealt{Mesinger2015, Mason2018a, Whitler2020, Wang2020}; Figure\ref{fig:xHI} and Table \ref{tab:T,xHI}).
    If this implies underestimation of our calculation due, for example, to an increase with redshift in the Ly$\alpha$ escape fraction of galaxies or the Ly$\alpha$ production rate per UV luminosity, then the lower limit to $\xHI(7.3)$ will also become higher than 0.28.
\end{enumerate}

\acknowledgments

We thank the anonymous referee for the constructive comments that greatly improved the manuscript.
We thank Rikako Ishimoto, Kei Ito, Ryohei Itoh, Ryota Kakuma, Shotaro Kikuchihara, Haruka Kusakabe, and Yongming Liang for useful comments and discussions.
KS is supported by the Toray Science Foundation.
TH is supported by Leading Initiative for Excellent Young Researchers, MEXT, Japan (HJH02007) and JSPS KAKENHI Grant number 20K22358.
R.M. acknowledges a Japan Society for the Promotion of Science (JSPS)
Fellowship at Japan and JSPS KAKENHI grant No. JP18J40088.
This work is supported by JSPS
KAKENHI Grant Number 17H01114 (AKI, SY, KS, and MO).

The Hyper Suprime-Cam (HSC) collaboration includes the astronomical communities of Japan and Taiwan, and Princeton University.  The HSC instrumentation and software were developed by the National Astronomical Observatory of Japan (NAOJ), the Kavli Institute for the Physics and Mathematics of the Universe (Kavli IPMU), the University of Tokyo, the High Energy Accelerator Research Organization (KEK), the Academia Sinica Institute for Astronomy and Astrophysics in Taiwan (ASIAA), and Princeton University.  Funding was contributed by the FIRST program from the Japanese Cabinet Office, the Ministry of Education, Culture, Sports, Science and Technology (MEXT), the Japan Society for the Promotion of Science (JSPS), Japan Science and Technology Agency  (JST), the Toray Science  Foundation, NAOJ, Kavli IPMU, KEK, ASIAA, and Princeton University.

This paper makes use of software developed for the Large Synoptic Survey Telescope. We thank the LSST Project for making their code available as free software at ${\langle}$http://dm.lsst.org${\rangle}$.

This paper is based on data collected at the Subaru Telescope and retrieved from the HSC data archive system, which is operated by Subaru Telescope and Astronomy Data Center (ADC) at NAOJ. Data analysis was in part carried out with the cooperation of Center for Computational Astrophysics (CfCA), NAOJ.

The Pan-STARRS1 Surveys (PS1) and the PS1 public science archive have been made possible through contributions by the Institute for Astronomy, the University of Hawaii, the Pan-STARRS Project Office, the Max Planck Society and its participating institutes, the Max Planck Institute for Astronomy, Heidelberg, and the Max Planck Institute for Extraterrestrial Physics, Garching, The Johns Hopkins University, Durham University, the University of Edinburgh, the Queen’s University Belfast, the Harvard-Smithsonian Center for Astrophysics, the Las Cumbres Observatory Global Telescope Network Incorporated, the National Central University of Taiwan, the Space Telescope Science Institute, the National Aeronautics and Space Administration under grant No. NNX08AR22G issued through the Planetary Science Division of the NASA Science Mission Directorate, the National Science Foundation grant No. AST-1238877, the University of Maryland, Eotvos Lorand University (ELTE), the Los Alamos National Laboratory, and the Gordon and Betty Moore Foundation.

\appendix
\section{Cutout images of spurious sources}
\label{appendix:A}
\begin{figure}[htb!]
    \centering
    \includegraphics[width=\linewidth]{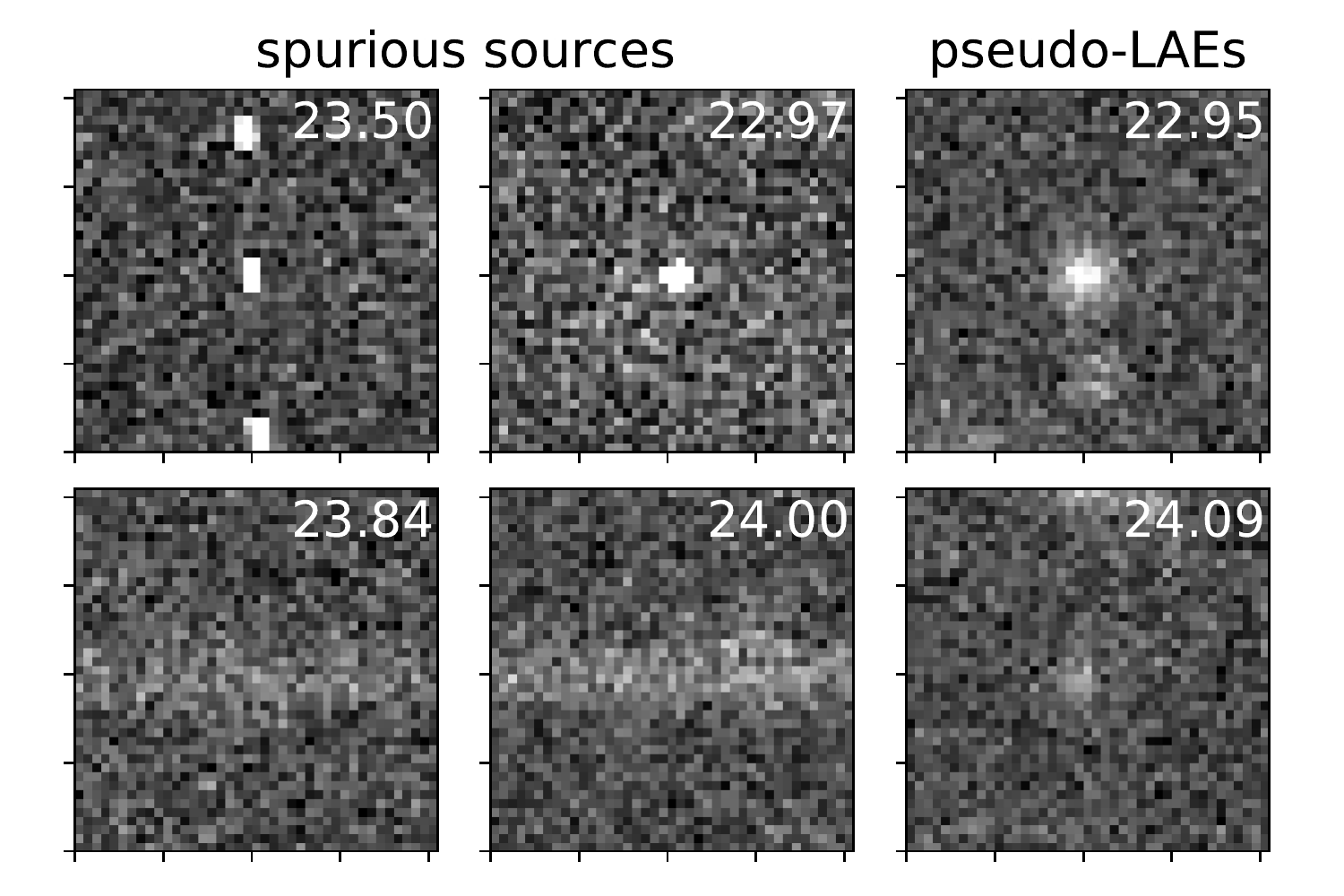}
    \caption{NB1010 images of four spurious sources (Section \ref{sec:LAEselection}; left two columns) and two pseudo-LAEs (Section \ref{sec:comp}; rightmost column). The numbers in the images are the aperture magnitudes, {\tt MAG\_APER}, of the objects. The size of each image is $6.\carcsec7 \times 6.\carcsec7$. North is up and east is to the left.}
    \label{fig:spurious}
\end{figure}

In Figure \ref{fig:spurious}, we present example images of spurious sources removed in our visual inspection (Section \ref{sec:LAEselection}).
For comparison,  we also show example images of pseudo-LAEs used in the calculation of completeness (Section \ref{sec:comp}).
The two spurious sources in the top row of this Figure are either a cosmic ray or a CCD artifact. Both have a well-outlined and highly-concentrated light distribution, with their total luminosity being contributed by only a small number ($\lesssim 10$) of pixels. The two spurious sources in the bottom row are diffuse, elongated structures. Each of them is due to a different, very bright star, and is located outside the mask for the star.

\section{Underestimation of $\TLya$ by the method using luminosity densities}
\label{appendix:B}
To obtain the relative transmission at a certain redshift, $\TLya(z)/\TLya(5.7)$,
we assume that observed Ly$\alpha$ luminosities are uniformly decreased in proportion to the IGM transmission (Equation (\ref{eq:LLya-LUV})). We then estimate this luminosity decrease by directly comparing the observed and predicted Ly$\alpha$ luminosities at a fixed cumulative number density (Equation (\ref{eq:T_L})).
Previous studies have, however, estimated $\TLya(z)/\TLya(5.7)$ from a decrease in the Ly$\alpha$ luminosity density as:
\begin{eqnarray}
\frac{\TLya(z)}{\TLya(5.7)} = \frac{\rho_{\Lya}(z)/\rho_{\Lya}(5.7)}{\rho_{\mathrm{UV}}(z)/\rho_{\mathrm{UV}}(5.7)},
\label{eq:T_rho}
\end{eqnarray}
where $\rho_{\Lya}$ and $\rho_{\mathrm{UV}}$ are the Ly$\alpha$ and UV luminosity densities, respectively, 
calculated by integrating the corresponding LFs.
Comparison with Equation (\ref{eq:T_L}) finds that $\rho_{\Lya}(z)$ is used in place of $L^\mathrm{obs}_{\Lya}(z)$ in Equation (\ref{eq:T_L}), and  $\rho_{\Lya}(5.7) \rho_{\mathrm{UV}}(z)/\rho_{\mathrm{UV}}(5.7)$ in $L^\mathrm{pred}_{\Lya}(z)$.
However, Equation (\ref{eq:T_rho}) is 
correct only when all four corresponding LFs are integrated down to zero luminosity.
Specifically, Equation (\ref{eq:T_rho}) overestimates the Ly$\alpha$ luminosity decrease and hence underestimates $\TLya(z)/\TLya(5.7)$ if the integration range of the two Ly$\alpha$ LFs is limited, as has been done by most previous studies: e.g., $\log L_{\Lya}\ [\ergs] =$ 42.4--44 has been adopted by \cite{Konno2018}, \cite{Itoh2018}, and \cite{Hu2019}.

To see why Equation (\ref{eq:T_rho}) overestimates the luminosity decrease with limited integration ranges of the Ly$\alpha$ LFs, let us assume a simple case that
Ly$\alpha$ luminosities are uniformly decreased due to IGM absorption
(the same assumption as in this study), and
the UV LF does not evolve with redshift.
In this case, the right hand side of Equation (\ref{eq:T_rho}) is reduced to $\rho_{\Lya}(z)/\rho_{\Lya}(5.7)$. 
If the Ly$\alpha$ LF at $z=5.7$ has the characteristic luminosity $L^*(5.7)$, then that at a redshift before completion of reionization will have $L^*(z) = L^*(5.7) \times \TLya(z)/\TLya(5.7)$, with the remaining two Schechter parameters \citep{Schechter1976}, $\phi^*$ and $\alpha$, being the same as of the $z=5.7$ LF because we have assumed a uniform luminosity decrease due to IGM absorption.
The question now is whether the equation $\rho_{\Lya}(z)/\rho_{\Lya}(5.7) = L^*(z)/L^*(5.7)$ is correct.
Using the Schechter parameters above, $\rho_{\Lya}(z)/\rho_{\Lya}(5.7)$ is written as:
\begin{eqnarray}
\frac{\rho_{\Lya}(z)}{\rho_{\Lya}(5.7)}
&=& \frac{\int_{L_{\mathrm{lim}}}^\infty L \phi(L^*(z),\phi^*,\alpha;L)dL}{\int_{L_{\mathrm{lim}}}^\infty L \phi(L^*(5.7),\phi^*,\alpha;L)dL} \nonumber \\
&=& \frac{ \int_{L_{\mathrm{lim}}}^\infty L\phi^* \left(\frac{L}{L^*(z)}\right)^\alpha\exp  \left(-\frac{L}{L^*(z)}\right) \frac{dL}{L^*(z)} }{  \int_{L_{\mathrm{lim}}}^\infty L\phi^* \left(\frac{L}{L^*(5.7)}\right)^\alpha\exp  \left(-\frac{L}{L^*(5.7)}\right) \frac{dL}{L^*(5.7)}} \nonumber \\
&=& \frac{L^*(z)}{L^*(5.7)} \times  \frac{\int_{L_{\mathrm{lim}}/L^*(z)}^\infty x^{\alpha+1} \exp(-x)dx}{\int_{L_{\mathrm{lim}}/L^*(5.7)}^\infty x^{\alpha+1} \exp(-x)dx}, \nonumber \\
&& 
\label{eq:rho_integral}
\end{eqnarray}
where $L_\mathrm{lim}$ is the integration limit and $x\equiv L/L^*(z)$.\footnote{We set the upper limit of the integral to infinity for simplicity, because the contribution from $\log L^* [\ergs]>44$ is negligible.}
The right hand side of the last line of this equation is lower than
$L^*(z)/L^*(5.7) (=\TLya(z)/\TLya(5.7))$,
because the integral at the numerator is smaller than that at the denominator owing to a narrower integration range of $x$ (because of $L_{\mathrm{lim}}/L^*(z) > L_{\mathrm{lim}}/L^*(5.7)$).

As an example, let us take $\log L^*\ [\ergs] = 43.2$ and $\alpha=-2.56$ at $z=5.7$ (the values obtained by \citealt{Konno2018}), and assume a $50\%$ luminosity decrease due to IGM absorption, i.e., $\TLya(z)/\TLya(5.7) = L^*(z)/L^*(5.7)=0.5$. In this case, Equation (\ref{eq:rho_integral}) with an integration range of $\log L_{\Lya}\ [\ergs] =$ 42.4--44 gives $\rho_{\Lya}(z)/\rho_{\Lya}(5.7)=0.23$, i.e., $77\%$ decrease.

On the other hand, \cite{Ouchi2010} and \cite{Konno2014}'s results include no systematic bias because they integrated the LF down to zero luminosity. However, their strategy instead leads to an extremely large uncertainty in the luminosity density due to a large extrapolation of the LF from the observed luminosity range.

\section{$\xHI$ estimates using other theoretical models}
\label{appendix:C}
Estimating $\xHI$ using the Ly$\alpha$ luminosity of galaxies requires a theoretical model that relates observed Ly$\alpha$ luminosities, or Ly$\alpha$ LFs, with $\xHI$.
To mitigate model dependence, we also estimate $\xHI$ using methods other than \cite{Jung2020}'s, as in previous studies.
First, using Ly$\alpha$ LFs for several $\xHI$ values simulated by \cite{Inoue2018} (their Figure 18; see also Section \ref{sec:comparison}), we obtain $\xHI(7.3)>0.2$ from our new data. 
Second, we use the analytic model of \cite{Santos2004} that calculates observed Ly$\alpha$ emission from an isolated galaxy at $z=6.5$ by varying many parameters associated with the galaxy and the IGM around it.
By comparing our $\TLya(z)/\TLya(5.7)$ estimates with their Figure 25, we obtain $\xHI$ for a galactic wind with a Ly$\alpha$ velocity offset of 0 and $360\ $ km s$^{-1}$, as shown in Table \ref{tab:Santos}.
These values are consistent with our results from the \cite{Jung2020} method within the errors.

\begin{deluxetable*}{clccc}[tb!]
\tablecaption{$\xHI$ estimates from \cite{Santos2004}\label{tab:Santos}}
\tablehead{\colhead{$z$} & \colhead{Ly$\alpha$ LF} & \colhead{$\TLya(z)/\TLya(5.7)$} & \multicolumn{2}{c}{$\xHI$}\\
&&& $v=0\ \kms$ & $v=360\ \kms$}
\startdata
7.3 & this study & $<0.77$ & $>0.2$ & $>0.2$\\
& \cite{Konno2014} & $0.53^{+0.18}_{-0.22}$ & $0.4^{+0.2}_{-0.2}$ & $0.5^{+0.3}_{-0.2}$\\
7.0 & \cite{Itoh2018} & $0.94^{+0.12}_{-0.17}$ & $0^{+0.2}$ & $0.1^{+0.1}_{-0.1}$\\
& \cite{Hu2019} & $0.90^{+0.11}_{-0.14}$ & $0.1^{+0.1}_{-0.1}$ & $0.1^{+0.1}_{-0.1}$\\
6.6 & \cite{Konno2018} & $0.95^{+0.14}_{-0.19}$ & $0^{+0.2}$ & $0.1^{+0.1}_{-0.1}$
\enddata
\end{deluxetable*}

\bibliography{main}{}
\bibliographystyle{aasjournal}
\end{document}